\documentclass[aip,graphicx]{revtex4-1}

\usepackage{amsmath}
\usepackage{graphicx,import}
\usepackage[usenames,dvipsnames]{xcolor}

\newcommand{\rev}[1]{\textcolor{Black}{#1}}
\newcommand{\revise}[1]{\textcolor{Black}{#1}}

\usepackage[normalem]{ulem}
\usepackage{subcaption}
\usepackage{mathtools}
\draft 

\begin{document}


\title{Mean flow data assimilation based on physics-informed neural networks} 



\author{Jakob G.~R.~von Saldern}
\email{j.vonsaldern@tu-berlin.de}
\affiliation{Laboratory for Flow Instabilities and Dynamics, Technische Universität Berlin, Berlin, 10623, Germany}

\author{Johann Moritz Reumschüssel}
\affiliation{Chair of Fluid Dynamics, Technische Universität Berlin, Berlin, 10623, Germany}
\author{Thomas L.~Kaiser}
\affiliation{Laboratory for Flow Instabilities and Dynamics, Technische Universität Berlin, Berlin, 10623, Germany}
\author{Moritz Sieber}
\affiliation{Laboratory for Flow Instabilities and Dynamics, Technische Universität Berlin, Berlin, 10623, Germany}
\author{Kilian Oberleithner}
\affiliation{Laboratory for Flow Instabilities and Dynamics, Technische Universität Berlin, Berlin, 10623, Germany}


\date{\today}

\begin{abstract}
Physics-informed neural networks (PINNs) can be used to solve partial differential equations (PDEs) and identify hidden variables by incorporating the governing equations into neural network training. In this study, we apply PINNs to the assimilation of \rev{turbulent} mean flow data and investigate the method's ability to identify inaccessible variables and closure terms from sparse data. Using high-fidelity large-eddy simulation (LES) data and particle image velocimetry (PIV) measured mean fields, we show that PINNs are suitable for simultaneously identifying multiple missing quantities in turbulent flows and providing continuous and differentiable mean fields consistent with the provided PDEs. \rev{In this way, consistent and complete mean states can be provided, which are essential for linearized mean field methods. The presented method does not require a grid or discretization scheme, is} easy to implement, and can be used for a wide range of applications, making it a very promising tool for mean field-based methods in fluid mechanics.
\end{abstract}


\maketitle 


\section{Introduction}
The determination of time-averaged flow fields is of great importance for many technical and scientific purposes. Such mean fields can be used, for example, for the calculation of drag coefficients of vehicles, stationary forces on turbine blades, or for the determination of static stresses on machine components in general.~\cite{Corson2009} \rev{ Moreover, consistent mean fields and associated closure fields are the input for linearized mean field analysis to study and predict dynamic flow phenomena such as global instabilities or dominant coherent structures.~\cite{barkley2006,Crouch2007,oberleithner2011,Beneddine2016} By consistency, it is meant that the mean fields satisfy the set of equations from which the linearized equations are derived; these are typically the Reynolds-averaged Navier-Stokes (RANS) equations.} 

When determining mean flows, one usually resorts to RANS simulations.~\cite{pope2000} An associated problem, however, is that this approach requires case-dependent closure models since large and medium scale coherent structures are domain and boundary condition dependent.~\cite{alfonsi2009} Thus, mean flows determined by RANS simulations often have insufficient accuracy, especially for technically relevant configurations. If this inaccuracy is not to be accepted, other, more elaborate methods must be employed to determine the mean flow. Very accurate mean fields can be obtained from high-fidelity simulations such as large-eddy simulations (LES) or from elaborate experimental measurements. Nevertheless, these methods also have drawbacks. In particular, measurement data are typically limited in spatial coverage. Experimental determination of pressure and temperature is usually limited to a few probe positions, and optical measurement techniques such as particle image velocimetry (PIV) are generally restricted to 2D sections of the 3D domain.~\cite{tropea2007} Other optical techniques are even further limited in that only integrated line-of-sight data can be recorded, such as the background-oriented schlieren (BOS) technique used to measure density fields.~\cite{raffel2015} Furthermore, such methods can only be applied if optical access to the relevant regions of the domain is given, which is problematic for many technical systems. 

High-fidelity simulations such as LES or direct numerical simulations (DNS) are not limited by such spatial constraints, since all flow variables can be resolved on the entire simulation grid.~\cite{lesieur1996} However, the simulation of technically relevant, complex systems with good temporal and spatial resolution often exceeds the computational budget. This is particularly problematic if statistical quantities are to be recorded for which long time series are required. Another problem are unknown boundary conditions such as heat transfer properties and surface properties of walls as well as the degree of turbulence at inlets. For thermo-fluid dynamics processes involving chemical reactions, like combustion processes, appropriate models for kinetics and heat release are needed for LES calculations which further increases the complexity.~\cite{poinsot2005} Therefore, RANS simulations often remain the method of choice for mean field determination after all, especially for extensive parameter studies and thermo-fluid problems. 

\rev{Two particular difficulties arise when using high-fidelity mean fields as input to linearized methods. First, the linearized equations are usually based on a RANS framework, so that the high-fidelity mean fields are not consistent with the linearized equations. Second, as mentioned above, the high-fidelity mean fields are often incomplete; measurement data are limited to certain quantities, and LES fields lack RANS-type closure fields, for example. Both are serious problems that often limit the use of linearized methods.~\cite{casel2022,pickering2021,manoharan2015,kuhn2022}}

One way to overcome the disadvantages of sparse experimental data is to assimilate mean flow fields based on the RANS equations. In this way, highly accurate and at the same time well-resolved and complete mean fields, including RANS-type closure fields, can be determined if measured data is available or partially available. \rev{An appropriate choice of equations in the assimilation procedure, can ensure that the assimilated mean fields are consistent with a given linearized frame.} 

These methods are known as mean flow data assimilation or field inversion and have attracted much attention in the last decade. For example, Li et al.~adjust the parameters of the $k-\omega$ turbulence model with an optimization procedure to adapt RANS predictions for three canonical flow geometries to measured data.~\cite{LI2017} Since the adaptation of individual model constants allows little flexibility, other studies have examined spatially varying adaptations. Based on a variational formulation approach (adjoint based optimization) the two dimensional flow around an infinite cylinder at $\text{Re}=150$ is assimilated by Foures et al.~\cite{foures2014} using the Reynolds stress fields as control parameter. In a subsequent study by Symon et al. the adjoint based technique is successfully applied to assimilate the flow field around an airfoil at $\text{Re}=13500$ from measured PIV data.~\cite{symon2017} Franceschini et al. overcome difficulties at high Reynolds numbers, by including the Spalart-Allmaras turbulence model in addition to a body forcing term in the momentum equations that serves as the control parameter, to assimilate the mean flow field over a backward facing step at $\text{Re}=28275$.~\cite{franceschini2020} The study also examines an adaptation of the source term of the Spalart-Allmaras turbulence model as the control parameter. Since this approach is limited to the Boussinesq hypothesis, as opposed to a volume forcing term acting on the momentum equations, it is less suitable for most flows. However, when the measured data are limited to a few measurement points, adapting the Spalart-Allmaras model proves to be the more robust method.~\cite{franceschini2020} Modification of the Spalart-Allmaras source term based on a discrete and  continuous adjoint formulation has also been considered for several other flow geometries by Singh and Duraisamy~\cite{singh2016} and He et al.~\cite{he2018}, respectively. Parish and Duraisamy propose an additional step after identifying an improved closure field, using machine learning to relate the assimilated field to mean field quantities to directly identify closure models.~\cite{PARISH2016}

In recent years, a new method called physics-informed neural networks (PINNs) has been developed in the field of scientific computing.~\cite{raissi_1} The method takes advantage of automatic differentiation~\cite{baydin2018} to incorporate information about physics in the form of partial differential equations (PDEs) in the training loss function of a neural network. In the field of fluid mechanics, the PINN methodology has already been applied to a large number of configurations and is constantly being adapted and further developed.~\cite{raissi_1,NSFnets,raissi_2,cai2021,eivazi2021,xu2021} On the one hand, the method can be used to solve partial differential equations in the classical sense, based on boundary and initial conditions. In the PINN terminology this is called solving forward problems. On the other hand, the method can also be used to identify unknown quantities, which fulfill the specified differential equations, based on given data. This can be used for data assimilation or identification of hidden variables and is referred to as solving inverse problems. For the present work, the latter is of primary interest, as this study is concerned with mean flow assimilation and closure identification. In contrast to classical mean flow assimilation approaches based on adjoint optimization, solving inverse problems with PINNs has remarkable advantages. The method does not require a grid, a discretization scheme, or a CFD solver. In fact, no equations other than the governing flow equations need to be derived. The method only requires typical optimization procedures to train the neural network and compute the derivatives of the neural network outputs with respect to its inputs, which can be done very efficiently with automatic differentiation. All routines needed for this method are already implemented in common machine learning toolboxes. 

In pioneer work Raissi et al. demonstrate the capability of PINNs for solving forward flow problems by computing a flow past a cylinder,~\cite{raissi_1} and inverse flow problems by assimilating velocity and pressure fields from simulated flow visualization measurements mimicked by concentration images of a transported passive scalar.~\cite{raissi_2} Based on this, the PINN method has recently been extensively researched and applied to many different types of flow problems. For the sake of clarity, only a few studies of particular relevance are reported here, for a more comprehensive overview we refer to the review paper of Cai et al..~\cite{cai2021} Solving RANS equations in a forward way based on PINNs is demonstrated by Eivazi et al..~\cite{eivazi2021}
The identification of missing flow data considering steady and unsteady RANS data is considered by Xu et al..~\cite{xu2021} In identifying hidden variables or assimilating missing data, i.e., solving inverse problems, mainly synthetic data have been considered so far.~\cite{raissi_2,cai2021,xu2021} Some of these studies investigate the effects of artificially added noise. Real experimental data for data assimilation based on PINNs has been used in two recent studies. Wang et al. apply a PINN to improve the quality of experimental three-dimensional tomographic PIV measurements in a wake flow of a hemisphere.~\cite{Wang2022} The results are compared to DNS solutions and show the great potential of PINNs to augment spatially sparse and noisy data into fully resolved three-dimensional flow fields.  
Cai et al. assimilate 3D velocity and pressure fields \rev{of a flow over an espresso cup} from tomographic BOS measurements based on the PINN methodology.~\cite{cai2021_espresso} \rev{Based on this, Molnar et al.~have developed the physics-informed background-oriented schlieren methodology, which allows the determination of velocity, density and pressure fields in supersonic flows based on BOS data.~\cite{Molnar22}} \rev{Post et al. consider PINNs for forward and inverse problems of time-averaged inviscid flows.~\cite{post2022} Their analysis includes the determination of the isentropic outlet Mach number of a turbine vane cascade based solely on boundary conditions and surface pressure measurements. These studies are particular examples} of how PINNs can identify hidden flow variables, opening up new possibilities for experimental approaches. 

\rev{Most of the PINN studies reported here have focused on either laminar or inviscid flows; turbulent mean flows have received limited attention. All the existing studies that consider turbulent mean flows have in common that the underlying equations of the neural network are consistent with the equations that were used to generate the training data under consideration, i.e., training RANS-informed networks with RANS data. In contrast, in this study we use high-fidelity LES and PIV mean fields to train networks which are informed by simplified models or reduced sets of equations. Herein, we consider a set of equations to be a simplified model whenever the applied mean fields do not necessarily satisfy the given equations. Consideration of simplified models is elementary for the assimilation of mean flow data in fluid mechanics, since time-averaged flow equations contain closure terms for most relevant configurations. In particular turbulent closure models often rely on assumptions, such as the Boussinesq hypothesis and thus the simplified flow model can only approximate the high-fidelity mean flow. To our knowledge, this is the first study that considers the assimilation of high-fidelity data to simplified models using PINNs. In distinction to simplified models, we speak of reduced equation sets when only a few equations are used that represent only parts of the physics underlying the data.} 


In the following, we briefly review the theory of physics-informed neural networks and then consider data assimilation and closure identification based on reduced equation sets and simplified models. Three examples of increasing complexity are considered. 
We begin by assimilating the density field from LES velocity data of a reacting turbulent jet by considering the continuity equation. In this first example, we are thus concerned with assimilation using reduced sets of equations rather than simplified equations, since the continuity equation is satisfied by the velocity data but only represent a part of the underlying physics. \rev{For a linearized analysis of reacting flows, the density field is essential.~\cite{casel2022}}
In contrast, we consider data assimiliation from a simplified model in the second case. Therefore, we assimilate the weakly non-parallel mean flow of an isothermal round turbulent jet using the boundary layer equation and LES velocity data. This increases complexity since the boundary layer equation do not represent the LES data exactly.
Finally, we consider the most typical simplified model in fluid dynamics, namely the RANS equations with a Boussinesq type eddy viscosity model. Based on this simplified model we assimilate the strongly non-parallel mean flow from measured axial and radial PIV velocity data of a turbulent swirl flow and thereby simultaneously identify the time-averaged values of the pressure, eddy viscosity and azimuthal velocity field. \rev{This last case illustrates how a consistent and complete set of mean fields can be generated from limited PIV measurement data, meeting all the requirements for a linearized mean field analysis. }

\section{Physics-Informed Neural Networks}\label{sec:PINN}
In this section, the basic idea of PINNs is outlined. For a more detailed introduction to the methodology, we refer to the original work by Raissi et al..~\cite{raissi_1} Although the methodology is also applicable to dynamic problems, we restrict ourselves herein to stationary problems, since the focus of the present study is mean flow data assimilation. Let $\bar{\mathbf{y}}$ be a set of mean flow field quantities (e.g.~velocities, pressure, eddy viscosity etc.) and let $\mathbf{x}$ denote the spatial coordinates. Then a set of stationary flow PDEs can be formulated as
\begin{align}
    \mathbf{0} = \underbrace{\mathcal{N}\left(\bar{\mathbf{y}}(\mathbf{x}),\frac{\partial \bar{\mathbf{y}}(\mathbf{x})}{\partial \mathbf{x}},\frac{\partial^2\bar{\mathbf{y}}(\mathbf{x})}{ \partial \mathbf{x}^2},... \right)}_{L_{\text{PDE}}}, \label{eq:N}
\end{align}
where $\mathcal{N}$ is any linear or nonlinear operator. The objective of data assimilation or solving inverse problems is to find unknown (hidden) mean flow quantities that satisfy these PDEs, based on some known or partially known quantities in the domain. In the principle of PINNs, this is achieved by using deep neural networks (DNNs) as universal function approximators.~\cite{Hornik1989} In the context of mean flow data assimilation, the DNN approximates the mean flow quantities as a function that depends on the spatial coordinates. Expressing the DNN as $\mathbf{f_{\alpha}}$ and it's parameters as $\mathbf{\alpha}$, this reads
\begin{align}
    \bar{\mathbf{y}} = \mathbf{f}_{\alpha}(\mathbf{x}) \label{eq:DNN}.
\end{align}
When identification of hidden variables is the objective, the output of the DNN $\bar{\mathbf{y}}$ includes both mean flow variables for which training data are available and unknown quantities - the hidden variables. In the PINN methodology, the function $ \mathbf{f_{\alpha}}$ is learned by taking into account available training data and the physical constraints, making the neural network physics-informed.

The integration of physical constraints into the neural network framework is achieved by the method of automatic differentiation~\cite{baydin2018} which enables an efficient computation of the derivative of the outputs of a DNN with respect to its input parameters, i.e., it allows to compute $\partial \bar{\mathbf{y}} / \partial \mathbf{x}  $ or any other higher order spatial derivative. As the inputs of the DNN are the spatial coordinates and the outputs are mean flow quantities, this allows to check for their validity with respect to satisfying physical PDEs, such as Eq.~\ref{eq:DNN}.  
The consideration of known data and physical constraints is realized by creating a composite loss term that penalizes both the deviation from training data and the residual of the PDEs. The parameters of the PINN are then found by minimizing this loss function with common optimization algorithms,
\begin{align}
    \mathbf{\alpha} = \text{argmin} \left( w_1 \underbrace{||L_{\text{train}}||}_{||\bar{\mathsf{y}}^*- \bar{\mathsf{y}}||} + w_2  ||L_{\text{PDE}}||\right), \label{eq:totalcost}
\end{align}
where $\bar{\mathsf{y}}$ represents the quantities in $\bar{\mathbf{y}}$, for which training data $\bar{\mathsf{y}}^*$ are given and $||\cdot||$ denotes a norm. In the scope of the present study we use a squared $\text{L}^2$-norm. The training data can either come from measurements or simulations, or be explicitly specified, e.g. based on boundary conditions. The training loss $||L_{ \text{train} }||$ thus enforces the network to match its output with the available training data and the minimization of the residual of the PDE $||L_{\text{PDE}}||$ ensures that the network's output approximates the physical equations. The spatial points in which the latter is evaluated are referred to as collocation points in the following. To compensate for the influence of the number of collocation points $N_c$ and the number of available training data $N_t$, the terms are normalized accordingly (included in the norm). To control the influence of the two different loss terms, the weight coefficients $w_1$ and $w_2$ can be used. If the PINN is to be constrained with multiple PDEs, the individual equations can also be weighted individually. For the present study the weights are fixed to $1$, however the training data term $||L_{\text{train}}||$ is evaluated with normalized quantities, while the PDE is evaluated in physical units. This results in the physical constraint being weighted slightly more than the training data. In the context of the present work, this choice of weighting has proved successful for all cases considered, but it should be noted that, depending on the configuration considered, this may also lead to excessive or insufficient weighting of the physical boundary condition and thus to training failure. For more details on the choice of weights, we refer to the study by Wang et al.~which presents a dynamic weighting algorithm that improves training by avoiding difficulties related to numerical stiffness.~\cite{wang2021}   

PDEs in cylindrical coordinates often contain terms with the radial coordinate in the numerator, as Eq.~\ref{eq:conti}. In our experience, this can lead to significant numerical problems in determining the PINN parameters. We expect this to be a general problem in solving PDEs in cylindrical coordinates with PINNs. For all problems considered in this study, multiplying the PDEs by the radial coordinate $r$, solved the problem and significantly increased the training speed. Moreover, we would like to note that in cylindrical coordinates the radial coordinate $r$ itself appears in the PDEs. Therefore, the physical loss term consists not only of the outputs of the DNN and their derivatives, but also of its input. This is a difference to the consideration of PDEs in Cartesian coordinates, but it does not cause any further difficulties.

In the scope of the present study we use PINNs with 10 hidden layers, each with 30 neurons and $\tanh$-activation functions. The output layers are composed of linear activation functions.
In choosing the network architecture, we followed other PINN studies in fluid mechanics.~\cite{cai2021} Nevertheless, we argue that there is no upper bound on the network size and complexity. Since PINN training is based on the inclusion of physical PDEs, and if the corresponding PDE loss term has sufficient weight in the overall loss function, unphysical overfitting is avoided. Yet, it should be noted that the larger the network, the more costly the training. On the other hand, the mean fields may not be approximated well enough if the network is chosen too small. In our experience, however, this problem is directly noticeable by the fact that the training data are not well approximated either.

The location of the collocation and training points is chosen so that the flow field is well resolved. In general, the solutions presented below do not depend strongly on the total number of points. In fact, a direct influence is excluded by normalizing each term in Eq.~\ref{eq:totalcost} by the corresponding total number of points (included in the norm). However, they have an indirect influence via their spatial distribution, as this affects the weighting of individual regions. Care should therefore be taken to select sufficient points in regions of interest, i.e., regions with high gradients, near boundaries, and regions with pronounced structures.

The physical constraints (PDEs) depend on the considered problem and will be reported in the following subsections. The implementation is realized using the Python package Tensorflow,~\cite{tensorflow2015-whitepaper} which provides all objects, functions and methods for the design of PINNs. Inspired by other studies in the field,~\cite{eivazi2021,xu2021} the search for the PINN parameters $\alpha$, i.e. the solution of the Eq.~\ref{eq:totalcost}, which is also called PINN training, is performed in two steps. First, the Adam optimizer with mini-batch optimization is applied,~\cite{kingma2014} followed by fine-tuning based on the full-batch limited memory Broyden-Fletcher-Goldfarb-Shanoo (L-BFGS) method.~\cite{liu1989} 
In the context of the present work, the same training strategy with $240$ Adam iteration steps followed by $60000$ L-BFFS steps is applied for all considered problems. During the optimization with the Adam algorithm, the batch size and learning rate are adjusted in six steps as shown in Table~\ref{tab:my_label}.

Finally, we would like to stress that data assimilation with PINNs, i.e., solving inverse problems, has three particularly remarkable properties. First, the method does not require a grid or numerical dicretization, making addressing of inverse problems easy to implement. Switching from one problem to another is straightforward, even when the physical constraints are completely different. Second, once the PINN parameters that minimize the loss are found, $\mathbf{f}_{\alpha}(\mathbf{x})$ provides a continuous representation of the mean flow quantities that can be evaluated at any point in space without the need for a discretization scheme. Moreover, PINNs are differentiable, and using automatic differentiation, the derivatives of the approximated mean flow quantities can be computed very efficiently. Third, the method allows to address problems that are difficult to be solved with common techniques. For example, if boundary conditions are noisy, or only partially known, the PINN method still yields accurate results, as shown below.

\begin{table}[t]
\caption{\label{tab:my_label}\rev{Number of iterations, batch size and learning rate used for PINN training with the ADAM algorithm.}}
\begin{ruledtabular}
    \begin{tabular}{lcccccc}
         Number of iterations\hspace{5mm} & 5 & 10 & 25 & 50 & 50 & 100  \\
         Batch size & $2^5$ & $2^5$ & $2^6$ & $2^7$ & $2^8$ & $2^9$ \\
         Learning rate & $10^{-2}$ & $5 \times 10^{-3}$ & $10^{-3}$ & $10^{-4}$ & $10^{-5}$ & $10^{-6}$
    \end{tabular}
\end{ruledtabular}
\end{table}

\section{Results}
In the following three sections, we consider inverse fluid mechanics problems of increasing complexity. We begin with the extraction of the density field $\bar{\rho}$ from LES mean fields of a reacting turbulent jet by only considering the physical constraint of mass conservation. Following this, we consider the assimilation of mean fields, using the turbulent boundary layer equation, from high-fidelity LES mean fields of a turbulent round jet. Here, the dominant component of the Reynolds stress tensor $\tau$ is treated as a hidden variable and is identified in the process. In the final case, we increase both the complexity in terms of the training data, using real experimental data, and in terms of the hidden variables to be identified. We apply the PINN methodology to assimilate the strongly non-parallel mean flow, including all three velocity components $\bar{u}_x \, \bar{u}_r \, \bar{u}_\theta$, pressure $\bar{p}$, and eddy viscosity $\nu_T$, from axial and radial PIV velocity fields of a turbulent swirling flow and the RANS equations. Table~\ref{tab:cases} summarizes the three cases considered, including their specific characteristics.

\begin{table}[ht]
    \caption{\rev{Summary of the considered cases.}}
    \begin{ruledtabular}
    \centering
       \begin{tabular}{lccc}
              & 1) reacting jet  & 2) isothermal jet & 3) swirling jet \\
             \hline
             PDE & continuity & boundary layer & RANS \\
             Training data & $\bar{u}_x \,\, \bar{u}_r$ & $\bar{u}_x \,\, \bar{u}_r$ & $\bar{u}_x \,\, \bar{u}_r \,\, (\bar{u}_\theta)$ \\
             Hidden variables & $\bar{\rho}$ & $\tau$ & $\nu_T \,\, \bar{p} \,\, \bar{u}_\theta$ \\
             Reduced equations & yes & yes & no \\
             Simplified equations \hspace{5mm} & no & yes & yes
        \end{tabular}
    \end{ruledtabular}
    \label{tab:cases}
\end{table}

\subsection{The turbulent reacting jet - density assimilation from velocity fields}
As a first example we consider a turbulent reacting jet at a Reynolds number of $\text{Re}=8000$ based on the nozzle diameter and the molecular viscosity as investigated before by Zhang et al..~\cite{zhang2021} Jet flames are of elementary importance for many technical devices. Since they have high axial velocity, this technology is particularly useful for flame stabilization of highly reactive fuels, such as hydrogen, which exhibit high flame speeds.~\cite{KALANTARI2017249} In order to experimentally extract high-precision field quantities in reactive jets, elaborate measurement techniques must be employed. Specifically, PIV measurements are commonly used to measure the velocity fields.
\rev{Other optical measurement methods provide insight into species concentration, temperature and pressure of the flow by measuring fluorescence or scattering of certain molecules.~\cite{Meier2002, hardalupas2004local} However, these methods often rely on complicated optical arrangements and/or complicated post-processing.} \rev{Another way to determine further quantities besides additional measurements} is to assimilate them from measured velocity fields.~\cite{symon2017}

\rev{Since the consideration of the density field is elementary to many analyses of reactive flows, such as linearized mean field analysis~\cite{manoharan2014,Terhaar2014b,Oberleithner2015a,Emerson2016,casel2022}, its assimilation from velocity data is considered in this first example.}
The density field can be computed by solving the continuity equation if the velocity fields are known. 
For an axis-symmetric flow the time averaged continuity equation reads
\begin{align}
    \underbrace{\bar{\rho} \left[ \frac{\partial \bar{u}_x}{\partial x} +\frac{\partial \bar{u}_r}{\partial r} + \frac{\bar{u}_r}{r} \right] + \bar{u}_x \frac{\partial \bar{\rho}}{\partial x} + \bar{u}_r \frac{\partial \bar{\rho}}{\partial r}}_{\mathcal{N}_\rho(\bar{u}_x,\bar{u}_r,\bar{\rho})} = 0,\label{eq:conti}
\end{align}
where $\bar{u}_x$ and $\bar{u}_r$ denote the mean axial and radial velocity component and $\bar{\rho}$ the mean density. Given the axial and radial velocity fields, Eq.~\ref{eq:conti} is a partial differential equation in $\bar{\rho}$ with non-constant coefficients, which can be solved with common numerical schemes. However, this approach entails two challenges. The first problem lies in noisy velocity measurements, which is especially problematic for computation of the spatial derivatives of $\bar{\mathbf{u}}$. Furthermore, the spatial resolution of the velocity fields are limited by the PIV measurement equipment which in turn limits the spatial resolution of the density fields to be calculated.
Both problems can be overcome if data assimilation is performed using the PINN methodology, where both the density and velocity fields are approximated as the output of a neural network. The goal of PINN training is to adjust the DNN's parameters such that the output of the network for $\bar{\mathbf{u}}$ approximates the measured velocity fields while also enforcing it to satisfying the continuity equation. 
As a result, one obtains both the velocity fields and the density field in the form of a continuous mapping. Thus, they can be resolved continuously in the entire domain and are additionally differentiable. The simultaneous step of fitting the velocity data and density identification allows the density field to be assimilated based on the continuity equation alone, without further preconditioning of the velocity fields.

To evaluate the applied PINN method, we use velocity fields of a turbulent jet flame generated by LES. Since the simulation also resolves the density, this allows the field assimilated with PINN to be validated directly. The LES data are obtained from Zhang et al..~\cite{zhang2021} The simulation was performed with openFOAM, where gas composition is modeled by a progress variable in combination with a chemistry lookup table. The reaction is taken into account by a turbulent flame speed closure model, also applied for other turbulent flame configurations by Zhang et al..~\cite{zhang2009,zhang2013,zhang2019} Since the focus of this paper is not on the LES, the reader interested in further details of the numerical strategy of the LES is kindly referred to the above cited literature.

For this application, the DNN maps the normalized spatial cylindrical coordinates to normalized axial and radial velocity fields, and to the density field,
\begin{align}
    \left[\bar{u}_x^\alpha(\tilde{x},\tilde{r}),\,\bar{u}_r^\alpha(\tilde{x},\tilde{r}),\,\rho^\alpha(\tilde{x},\tilde{r}) \right]^\text{T} = \mathbf{f}_{\alpha}(\tilde{x},\tilde{r}).
\end{align}
The use of bounded (normalized) inputs and outputs is advantageous because using physical quantities would require the DNN to map from/to different scales, which can result in training difficulties. Therefore, normalized coordinates are used, which are indicated by a tilde hereafter. \revise{Normalized PINN outputs and LES data are denoted with $\alpha$ and $*$ superscripts, respectively.}

\begin{figure*}[b!]
    \centering
    \includegraphics[width=\linewidth]{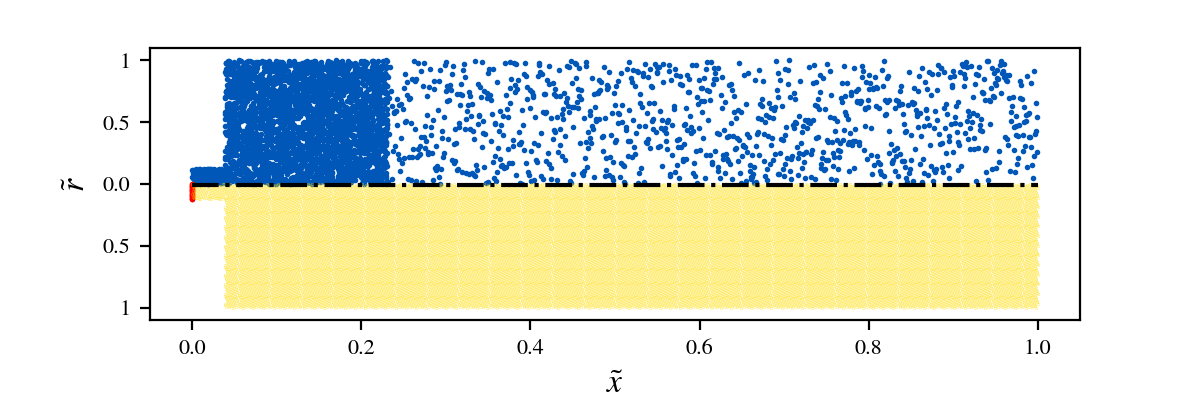}
    \caption{Computational domain of the reacting turbulent jet. Top: Blue dots show the collocation points. Bottom: Yellow dots show locations of points, in which velocity training data was provided to the PINN, red dots indicate the location at which a boundary condition for the density field is set. \revise{The axial and radial coordinate are normalized with the maximum domain lengths, $0.52$~m and $0.08$~m, respectively.}}\label{fig:turb_colloc_3f}
\end{figure*}

The composite loss function includes a training loss term and a PDE loss term, see Eq.~\ref{eq:totalcost}. To evaluate the training loss $N_t=20679$ normalized velocity training data points $\bar{u}_x^*(\tilde{\mathbf{x}}_i)$ and  $\bar{u}_r^*(\tilde{\mathbf{x}}_i)$ are sampled from the LES mean fields. The locations at which training data was provided are shown as yellow points in Fig.~\ref{fig:turb_colloc_3f}. To evaluate the physical loss term, $N_c=4200$ collocation points are selected within the domain.
The evaluation of the PDE at the collocation points then forms the physical loss, $||L_{\text{PDE}}||$ in Eq.~\ref{eq:totalcost}. The blue dots in Fig.~\ref{fig:turb_colloc_3f} show the collocation points used for this case. As the flame stabilizes close to the location of the nozzle, more collocation points are selected in this region in order to give special weight to compliance with physics in this domain. In total 3200 points are selected in the higher weighted nozzle region including 200 upstream of the area jump and 1000 points are selected in the less weighted downstream region. Within each region the collocation points are selected using Latin Hypercube Sampling. 

A closer look at Eq.~\ref{eq:conti} shows that the trivial solution $\bar{\rho}=0$ satisfies the PDE. In fact, for a given velocity field, there are infinitely many solutions for each streamline, since the PDE is linear. Thus, to formulate a well-posed problem, a boundary condition must be specified for each streamline. For this purpose, a Dirichlet boundary condition for the density field is imposed at the inlet. This is realized by an additional term in the cost function which penalizes deviations from $\bar{\rho}^\alpha=1$ at $N_b = 20$ locations at the inlet. The red points at the inlet ($\tilde{x} = 0$) in Fig.~\ref{fig:turb_colloc_3f} indicate the locations at which the density boundary condition is set. Although we apply the constraint at the inlet, we briefly note here that unlike conventional PDE solvers, for PINNs the boundary condition need not necessarily be applied at the boundary of the domain, it is sufficient if a density value for each streamline is given somewhere in the domain.

In summary, the individual terms of the composite loss function, Eq.~\ref{eq:totalcost} read
\begin{align}
\begin{split}
     ||L_{\text{train}}|| &=  \frac{1}{2N_t} \sum_{i=1}^{N_t}\left(\bar{u}_x^*(\tilde{\mathbf{x}}_i) - \bar{u}_x^\alpha(\tilde{\mathbf{x}}_i)\right)^2+\left(\bar{u}_r^*(\tilde{\mathbf{x}}_i) - \bar{u}_r^\alpha(\tilde{\mathbf{x}}_i)\right)^2 + ... \\
     &\quad  \frac{1}{N_b}\sum_{j=1}^{N_b} \left(1-\rho^\alpha(0,\tilde{r}_i)\right)^2,
     \end{split}\\
     ||L_{\text{PDE}}|| &= \frac{1}{N_c}\sum_{i=1}^{N_c} \left(r\mathcal{N}_\rho\left(U_x\bar{u}_x^\alpha(\tilde{\mathbf{x}}_i),\,U_r\bar{u}_r^\alpha(\tilde{\mathbf{x}}_i),\,R\rho^\alpha(\tilde{\mathbf{x}}_i)\right)\right)^2,
\end{align}
where the $\alpha$ and $*$ superscripts denote \revise{normalized} PINN and LES mean flow quantities, respectively. As discussed in Section~\ref{sec:PINN}, the training data term is evaluated with normalized quantities and the PDE is evaluated in physical units, where \revise{$U_x=5.95$~m/s and $U_r=2.03$~m/s} denote the maximum axial and radial velocity and $R$ denotes the maximum density. The maximum value of the density is unknown, however, since the PDE is linear, any non-zero value $R$ can be used. Since $R$ directly weights the physical loss term of the cost function, see Eq.~\ref{eq:conti}, we used the value of the unburned gas \revise{1.11}~kg/m$^3$ to avoid a further change in weighting. The PDE is multiplied by the radial coordinate to avoid numerical problems and to increase convergence speed.

\begin{figure*}[t!]
    \centering
    \begin{subfigure}{\linewidth}
        \includegraphics[width=0.96\linewidth]{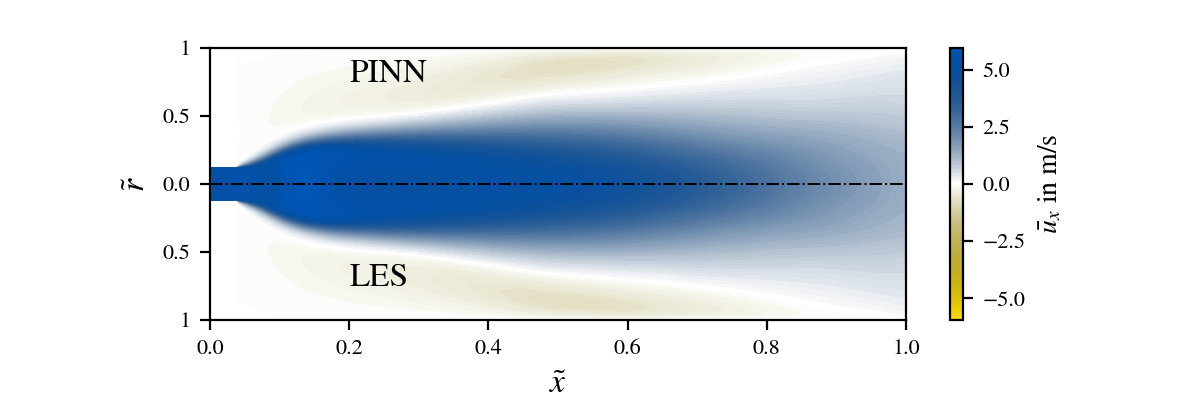}
        \caption{\revise{PINN and LES axial velocity}}
    \end{subfigure}
    \begin{subfigure}{\linewidth}
        \includegraphics[width=0.96\linewidth]{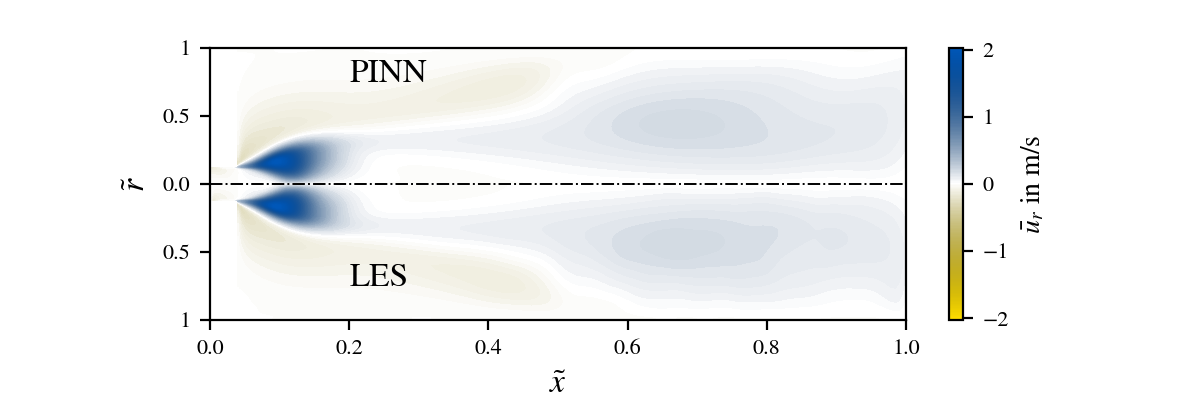}
        \caption{\revise{PINN and LES radial velocity}}
    \end{subfigure}
    \begin{subfigure}{\linewidth}
        \includegraphics[width=0.96\linewidth]{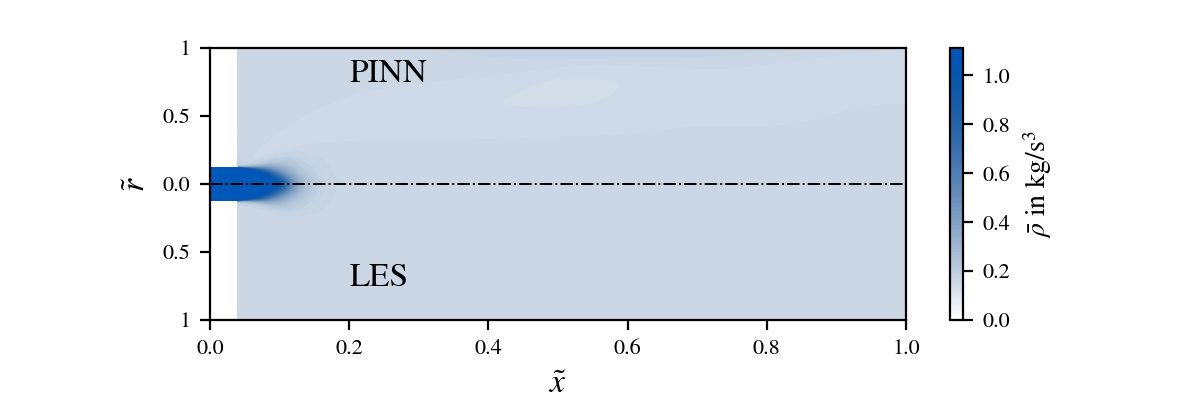}
        \caption{\revise{PINN and LES density}}
    \end{subfigure}
    \caption{Mean fields of the reacting jet. (a) axial velocity, (b) radial velocity and (c) density. Upper halves show the PINN results, lower halves show the respective LES validation data. \revise{The axial and radial coordinate are normalized with the maximum domain lengths, $0.52$~m and $0.08$~m, respectively.}}\label{fig:turb_4f}
\end{figure*} 

\begin{figure*}[b]
    \centering
    \begin{subfigure}{0.25\linewidth}
    \includegraphics[width=\linewidth]{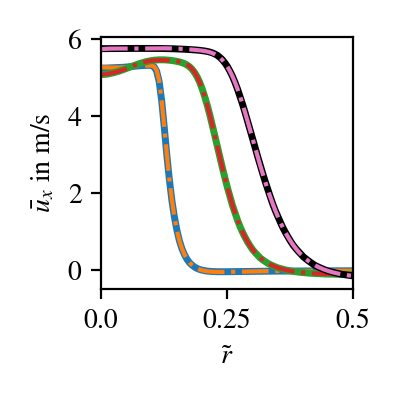}
        \caption{\revise{axial velocity}}\label{fig:turb_line_4fa}
    \end{subfigure}
    \begin{subfigure}{0.25\linewidth}
    \includegraphics[width=\linewidth]{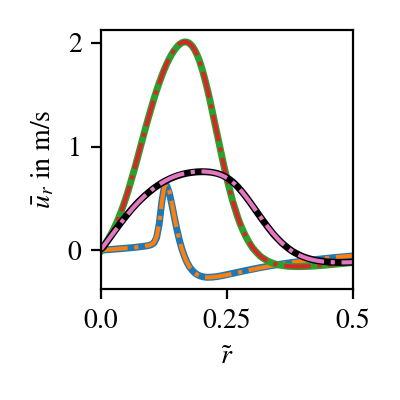}
        \caption{\revise{radial velocity}}\label{fig:turb_line_4fb}
    \end{subfigure}
    \begin{subfigure}{0.43\linewidth}
    \includegraphics[width=\linewidth]{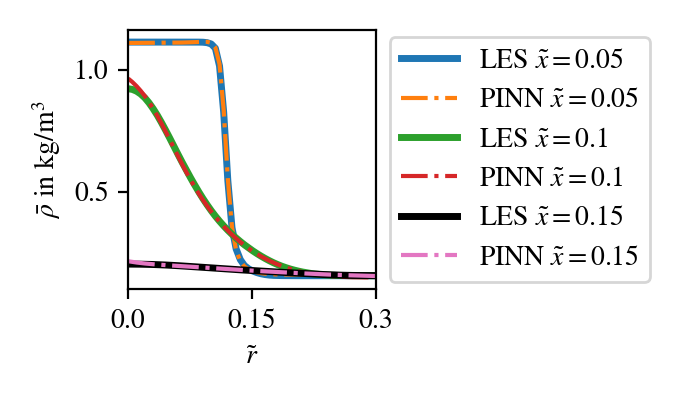}
        \caption{\revise{density}}\label{fig:turb_line_4fc}
    \end{subfigure}
    \caption{\revise{Mean field profiles at three axial positions ($\tilde{x}=0.05$, $\tilde{x}=0.1$, $\tilde{x}=0.15$). (a) axial velocity, (b) radial velocity, (c) density.} Dashed and solid lines show LES and PINN quantities, respectively. \revise{The axial and radial coordinate are normalized with the maximum domain lengths, $0.52$~m and $0.08$~m, respectively.}}\label{fig:turb_line_4f}
\end{figure*} 

The parameters of the PINN are found by minimizing the composite loss function, with the Adam and L-BFGS algorithms, as explained in Section~\ref{sec:PINN}. \rev{Fig.}~\ref{fig:turb_4f} shows the axial and radial velocity fields and the density field from top to bottom, with the lower halves showing the LES fields. The flow resembles that of a non-reacting round jet, with the fanning out of the jet being enhanced by the strong local heat release of the flame and the associated thermal expansion. The position of the flame can best be determined from the density field. Along the flame front, the density gradient is maximum due to the rapid temperature change. The upper halves in Fig.~\ref{fig:turb_4f} show the corresponding PINN solutions for the velocity components and the density. Additional line plots at three different exemplary axial positions are shown in Fig.~\ref{fig:turb_line_4f}, for a more thorough comparison between PINN and LES. One location ($\tilde{x} = 0.05$, Fig.~\ref{fig:turb_line_4fa}) is chosen near the nozzle exit, where the flame is located and a steep density gradient is present due to the strong temperature increase, and two axial locations slightly further downstream ($\tilde{x} = 0.1$, Fig.~\ref{fig:turb_line_4fb} and $\tilde{x} = 0.15$, Fig.~\ref{fig:turb_line_4fc}) are shown where the density gradient is less steep but still recognizable.
Overall, both the velocity fields and the assimilated density field match the LES data closely over the entire range.

\begin{figure*}[t]
    \centering
    \includegraphics[width=\linewidth]{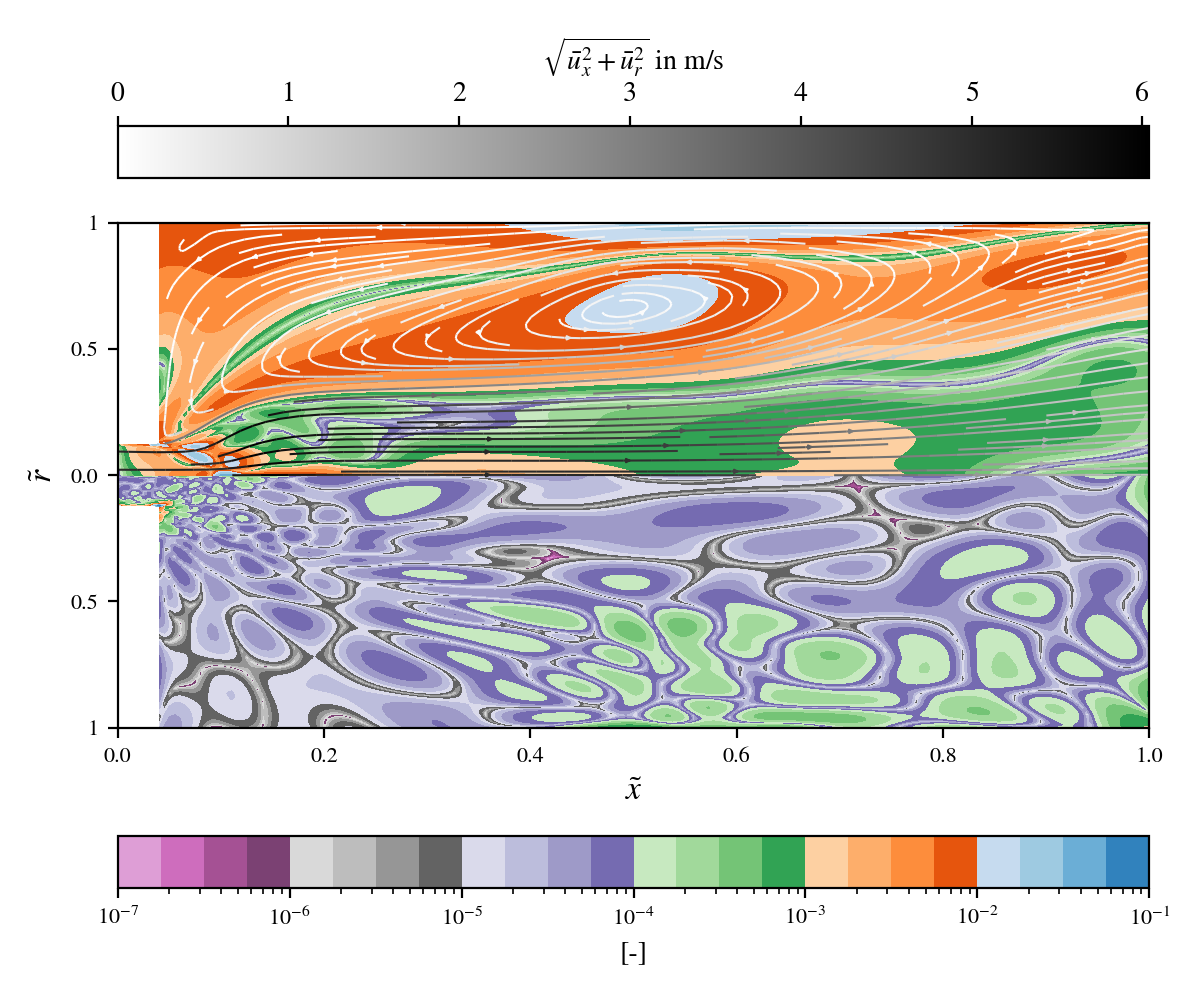}
    \caption{Density identification error analysis. Top: Normalized validation error $E_\rho$ and streamlines of the flow. Bottom: Normalized PDE residual $e_\rho$. The color bar at the top indicates the magnitude of the velocity along the streamlines. The color bar at the bottom shows the magnitude of the validation error and PDE residual. \revise{The axial and radial coordinate are normalized with the maximum domain lengths, $0.52$~m and $0.08$~m, respectively.}}\label{fig:turb_res_4f}
\end{figure*}

\rev{For a quantitative evaluation of the result, Fig.~\ref{fig:turb_res_4f} shows the absolute value of the normalized residual of the PDE, $e_\rho \coloneqq |r\mathcal{N}_\rho/(RU_x)|$, and the relative validation error of the density, \revise{$E_\rho \coloneqq | \bar{ \rho }^\alpha-\bar{\rho}^* |$. Note that $\bar{ \rho }^\alpha$ and $\bar{ \rho }^*$ are normalized quantities.} The two errors are evaluated at $N_e=38600$ points on an equidistant grid. The corresponding fields of $E_\rho$ and $e_\rho$ are shown in the upper and lower halves of Fig.~\ref{fig:turb_res_4f}, respectively. Note that $N_e$ exceeds the number of collocation points $N_c$, at which information about the PDE is provided to the PINN, by a factor of about 10. Thus, we take advantage of the fact that the PINN is a continuous and differentiable function that can be evaluated at any point. Fig.~\ref{fig:turb_res_4f} additionally shows the streamlines in the upper half, colored according to the magnitude of the velocity. At this point, it should be reiterated that the PINN receives no direct information about the density except through the inlet boundary condition, and assimilates it within the domain based solely on the PDE.
}

\rev{The residuum of the PDE shown in the lower part of Fig.~\ref{fig:turb_res_4f} has a spatial mean value of $6.9\times 10^{-5}$. Only in a small area near the nozzle exit, where the geometry changes abruptly, higher values of the PDE residual are visible near the wall. Here $e_\rho$ reaches its maximum of $7.2\%$. In this region, where the flow field has a complex shape and exhibits strong gradients, the PINN does not satisfy the PDE particularly well at locations that were not explicitly considered during training. If $e_\rho$ is evaluated at the collocation points only, the maximum value is one order of magnitude lower. We thus suspect that considering additional collocation points in this region could significantly reduce the residual.}

\rev{The validation error $E_\rho$ shown in the upper part of Fig.~\ref{fig:turb_res_4f} has mean and maximum values of $0.33\%$ and $4.4\%$, respectively, based on the $N_e$ points. Detailed inspections reveal that the area where the residual is high is limited to the region close to the wall at the nozzle exit. The validation error, on the other hand, also takes on comparatively high values at points downstream of the nozzle in regions with low residual. We suspect several reasons for this. First, the local inaccuracy at the near-wall nozzle exit, which can also be seen in the PDE residual, propagates through the domain. However, in the downstream region, this remains undetected in the PDE residual because the deviating density propagates according to the underlying PDE. Second, in regions with low velocities and small gradients, the residual of the PDE is comparatively small for any density. This means that identification of the density field based on the velocity data in these regions is difficult without adjusted weighting of the corresponding points in the composite loss function. It is evident from the streamlines that near the wall, downstream of the nozzle, the velocities and their gradients are small, and a correspondingly higher error is found in this region. In addition, a larger deviation is observed in the center of the recirculation zone. However, an increased deviation in this region is not surprising since the streamlines in the recirculation zone do not intersect the location where the boundary condition for the density is given. It is therefore remarkable that the density in this region can still be estimated with such high accuracy.}

\rev{This leads to the following conclusions for the assimilation of the density field based on velocity data and the continuity equation: The residual of the PDE does not directly indicate the quality of the approximation, and a local peak in the residual can potentially affect the estimate of the density in the downstream region. In regions where velocities and gradients are low, assimilation also proves to be more difficult. Nevertheless, it should be noted that, as shown in this example, the PINN can already approximate the density field with very high accuracy without any further action. Actions that can be taken to further increase accuracy include adjusting collocation points based on the PDE residual, locally adjusted weights, or providing more information to the PINN, using boundary conditions or measurements within the domain.}

\subsection{The turbulent round jet -  mean flow assimilation with \rev{the turbulent boundary layer equations}}
Time averaging of flow equations for turbulent flows leads to closure terms that must be modeled in order to calculate mean flows based on the set of averaged equations. The best known example of closure terms are the Reynolds stresses that appear in the famous Reynolds-averaged Navier-Stokes equations. The most common modeling approach, is the Boussinesq-type eddy viscosity model which relates the Reynolds stresses to the mean flow gradient using a turbulent viscosity - the eddy viscosity.~\cite{SCHMITT2007} Since this modeling approach corresponds to a strong assumption for most flows, the resulting mean flow equations can be considered a simplified model.

To obtain accurate results from time-averaged equations, modeling of the closure terms is crucial. Consequently, assimilation of mean flows based on time-averaged equations and measured or simulated mean fields of turbulent flows also requires correct identification of the closure terms. Since these are usually based on simplified equations, it is of great importance for the evaluation of existing and the development of new closure models to be able to perform the assimilation on the basis of simplified equations. Recent studies have shown that RANS closure models are also of great importance for dynamic models based on linear equations, since the same closure terms appear in the linear equations as in the RANS equations if the equations are linearized around the temporal mean.~\cite{kaiser_2021} \rev{The state of the art for identifying closure models for linearized mean field methods is a least squares fit of the eddy viscosity using the velocity field, but in many cases this leads to poor results.} Assimilated closure fields can thus not only be used for the development of new closure models but also for dynamic predictions using linearized analysis.

The second example in this study therefore investigates whether the PINN method can also be used to adjust simplified mean field models to high-fidelity data and thereby assimilate compatible closure fields. For this purpose, we consider the mean velocity fields of a weakly non-parallel circular jet at a Reynolds number of $10000$. To obtain the fields, LES snapshots from a simulation with openFOAM were time-averaged. More details on the simulation can be found in the study of Casel et al..~\cite{casel2022} \rev{For assimilation}, the simplified equations of motion for circular jets are considered,~\cite{rajaratnam1976}
\begin{align}
    \frac{\partial \bar{u}_x}{\partial x} + \frac{\partial \bar{u}_r}{\partial r} + \frac{\bar{u}_r}{r} &= 0 \label{eq:ROM_conti} \\
    \underbrace{\bar{u}_x\frac{\partial \bar{u}_x}{\partial x} +  \bar{u}_r\frac{\partial \bar{u}_x}{\partial r} -\frac{1}{\bar{\rho}} \left(\frac{\tau}{r} + \frac{\partial \tau}{\partial r}\right)}_{\mathcal{N}_\tau(\bar{u}_x, \bar{u}_r,\tau)} &= 0 \label{eq:ROM_mom1} ,
\end{align}
where $\tau$ is the shear stress term that represents the dominant Reynolds stress term $\tau = -\bar{\rho}\overline{u_r'u_x'}$ for round jets. \rev{Dashed variables indicate fluctuation variables according to the Reynolds decomposition ($u=\bar{u}+u'$).} \rev{Eq.}~\ref{eq:ROM_conti} is the incompressible continuity equation and Eq.~\ref{eq:ROM_mom1} describes the dominant momentum transfer in turbulent round jets. The latter is derived by time-averaging the 2D incompressible Navier-Stokes equations, assuming $\bar{u}_x \gg \bar{u}_r$, \rev{$\partial /\partial r \gg \partial / \partial x$}, neglecting viscous stresses and assuming a zero pressure gradient.~\cite{rajaratnam1976} \rev{Eq.}~\ref{eq:ROM_mom1} is also known as the turbulent boundary layer equation.
By further applying the Boussinesq hypothesis,~\cite{SCHMITT2007} the shear stress term could be related to the radial derivative of the mean axial velocity component, $\tau = \bar{\rho}\nu_T \partial \bar{u}_x / \partial r$. However, since only one shear stress term appears in the equation, the Boussinesq model does not introduce any further restriction and the use of either $\tau$ or an eddy viscosity $\nu_T$ is equivalent.

\rev{Eqs.}~\ref{eq:ROM_conti} and \ref{eq:ROM_mom1} are simplified equations that approximate the mean flow field of circular jets. The set of equations is not closed but contains a generally unknown closure term, the shear stress term $\tau$ (or eddy viscosity $\nu_T$ if the Boussinesq approximation is applied). If the equations are to be solved in a forward way this term must be modeled with an appropriate closure model. Here, however, we are interested in solving the inverse problem of assimilating the term from the given velocity mean fields. This is achieved by adjusting the equations to the mean LES fields using the PINN method. For this purpose the DNN is defined as the mapping of the normalized spatial coordinates to the normalized mean axial and radial velocity field, and the normalized shear stress term, 
\begin{align}
    \left[\bar{u}_x^\alpha(\tilde{x},\tilde{r}),\,\bar{u}_r^\alpha(\tilde{x},\tilde{r}),\,\tau^\alpha(\tilde{x},\tilde{r}) \right]^\text{T} = \mathbf{f}_{\alpha}(\tilde{x},\tilde{r}).
\end{align}
The composite loss function, Eq.~\ref{eq:totalcost}, for this case includes a training loss term with respect to the given mean velocity fields and the physical loss term that penalizes the violation of the simplified model equation. The shear stress field can be assimilated directly based on the turbulent boundary layer equation (Eq.~\ref{eq:ROM_mom1}) alone, without using the mass balance, in fact, the term does not even appear in the continuity equation. Thus, this case involves both a simplified model and a reduced set of equations. The individual terms of the composite loss function read
\begin{align}
     ||L_{\text{train}}|| &=  \frac{1}{2N_t} \sum_{i=1}^{N_t}(\bar{u}_x^*(\tilde{\mathbf{x}}_i) - \bar{u}_x^\alpha(\tilde{\mathbf{x}}_i))^2+(\bar{u}_r^*(\tilde{\mathbf{x}}_i) - \bar{u}_r^\alpha(\tilde{\mathbf{x}}_i))^2, \\
     ||L_{\text{PDE}}|| &= \frac{1}{N_c}\sum_{i=1}^{N_c} (r\mathcal{N}_\tau(U_x\bar{u}_x^\alpha(\tilde{\mathbf{x}}_i),\,U_r\bar{u}_r^\alpha(\tilde{\mathbf{x}}_i),T\tau^\alpha(\tilde{\mathbf{x}}_i)))^2,
\end{align}
where the $\alpha$ and $*$ superscripts again denote PINN and LES mean flow quantities, respectively, and the PDE is multiplied with the radial coordinate to avoid numerical problems. As in the previous case, the loss term, which takes into account the physical constraint, is weighted slightly higher because the equation is considered in physical units, while the training data are normalized. Since no training data are used for the shear stress, the normalization factor $T$ must be estimated. We have roughly estimated this value to be 1 $\text{kg}/\text{m}\text{s}^2$. Other values were also tested (not shown), and the results presented below proved to be very robust with respect to this parameter. \revise{For the normalization of the velocities the respective maximum values from the LES are used, $U_x=5.28$~m/s and $U_r=0.17$~m/s.} It should be noted that although the mass balance does not contain a term to be identified, it could still be added as an additional constraint for fitting the velocity fields.

\begin{figure*}[t!]
    \centering
    \includegraphics[width=\linewidth]{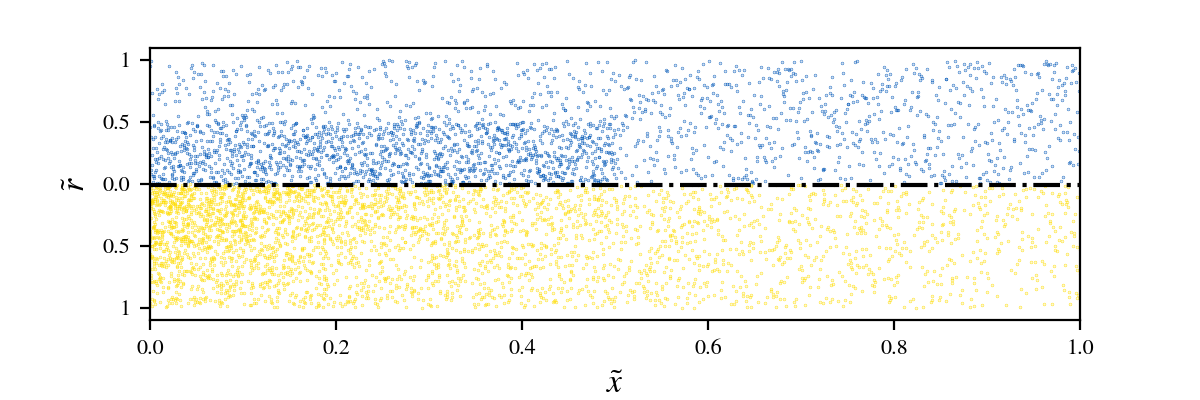}
    \caption{Computational domain of the turbulent jet. Top: Blue dots
    show collocation point at which the PDEs are evaluated. Bottom: Yellow dots show locations of points, in which velocity training data was provided to the PINN. \revise{The axial and radial coordinate are normalized with the maximum domain lengths, $0.88$~m and $0.20$~m, respectively.}}\label{fig:colloc_ROM_JFM}
\end{figure*} 
\begin{figure*}[t!]
    \centering
    \begin{subfigure}{\linewidth}
        \includegraphics[width=0.96\linewidth]{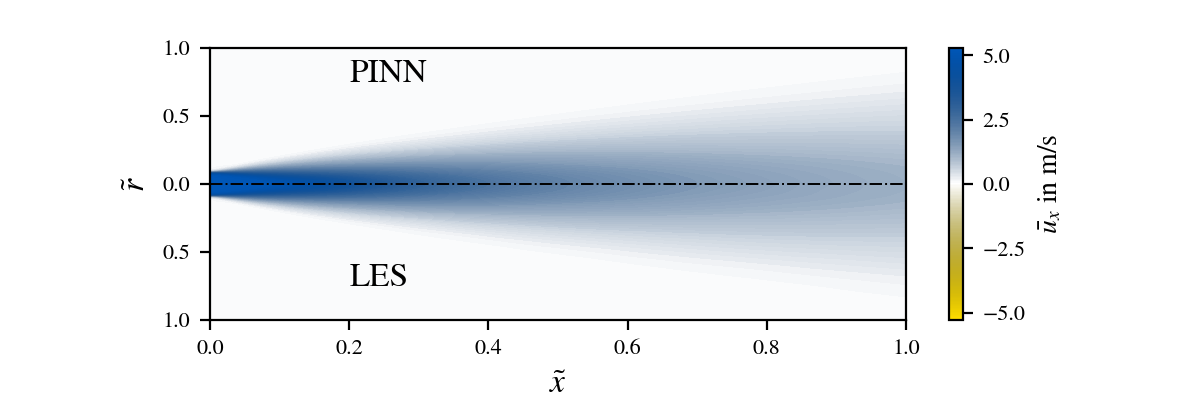}
        \caption{\revise{PINN and LES axial velocity}}\label{fig:ROM_JFMa}
    \end{subfigure}
    \begin{subfigure}{\linewidth}
        \includegraphics[width=0.96\linewidth]{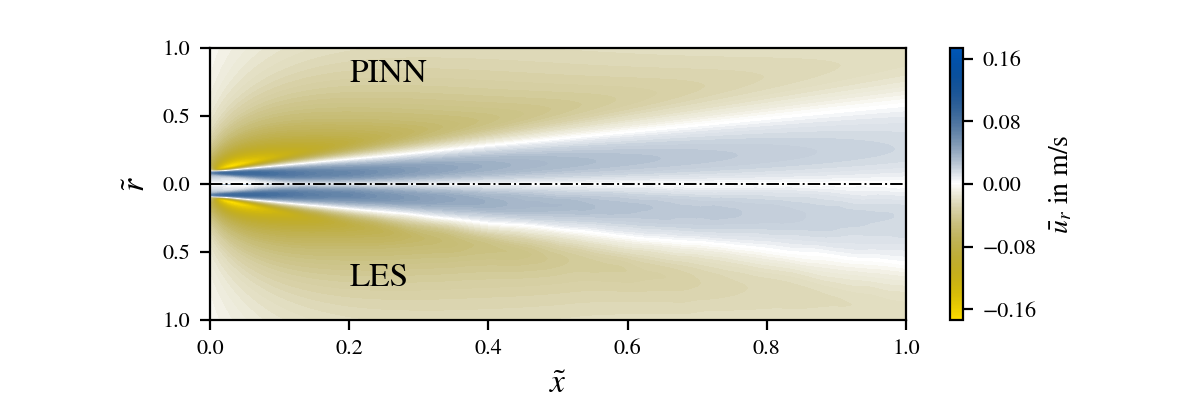}
        \caption{\revise{PINN and LES radial velocity}}\label{fig:ROM_JFMb}
    \end{subfigure}
    \begin{subfigure}{\linewidth}
        \includegraphics[width=0.96\linewidth]{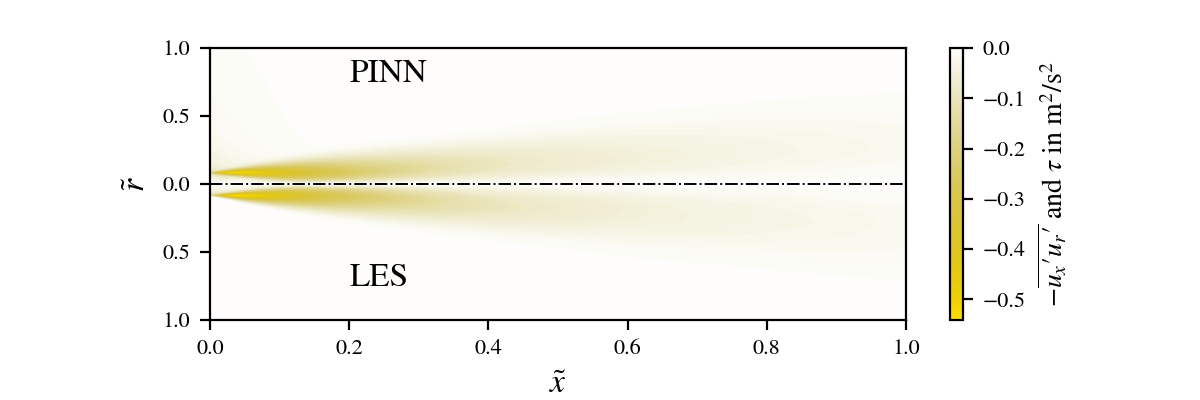}
        \caption{\revise{PINN and LES shear stress term}}\label{fig:ROM_JFMc}
    \end{subfigure}
    \caption{\revise{Mean} fields of the turbulent jet. (a) axial velocity, (b) radial velocity and (c) shear stress term. Upper halves show the PINN results, lower halves show the respective LES validation data. \revise{The axial and radial coordinate are normalized with the maximum domain lengths, $0.88$~m and $0.20$~m, respectively.}}\label{fig:ROM_JFM}
\end{figure*} 

\begin{figure*}[ht]
    \centering
    \begin{subfigure}{0.31\linewidth}
        \includegraphics[width=\linewidth]{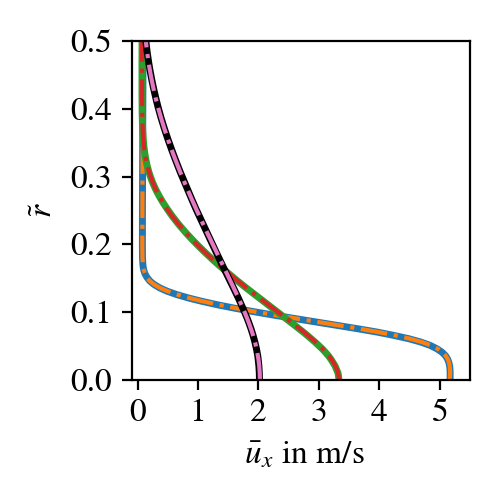}
        \caption{\revise{axial velocity}}\label{fig:ROM_JFM_lineA}
    \end{subfigure}
    \begin{subfigure}{0.31\linewidth}
        \includegraphics[width=\linewidth]{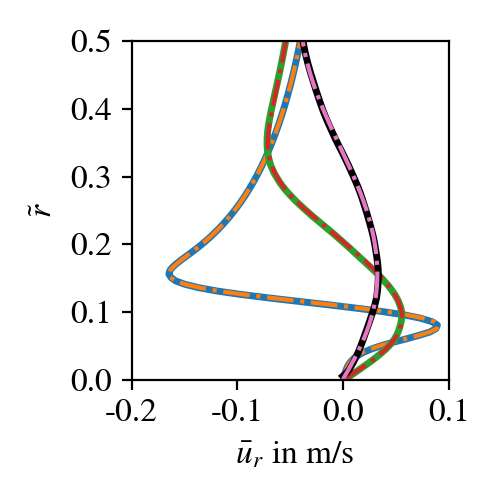}
        \caption{\revise{radial velocity}}\label{fig:ROM_JFM_line_ur}
    \end{subfigure}
    \begin{subfigure}{0.31\linewidth}
        \includegraphics[width=\linewidth]{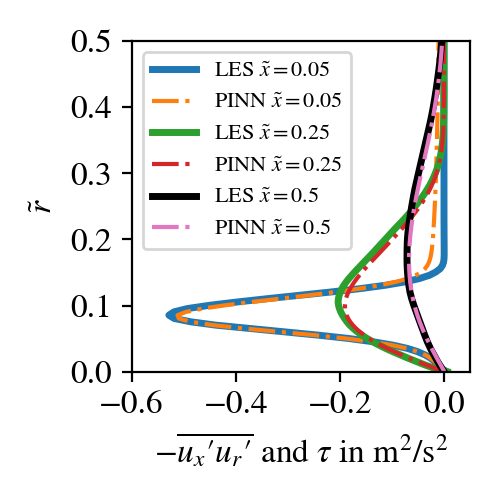}
        \caption{\revise{shear stress term}}\label{fig:ROM_JFM_lineB}
    \end{subfigure}
    \caption{\rev{Mean field profiles at three axial positions ($\tilde{x}=0.05$, $\tilde{x}=0.25$, $\tilde{x}=0.5$). \revise{(a) axial velocity, (b)} radial velocity and (c) shear stress term}. Dashed and solid lines show PINN and LES quantities, respectively. \revise{The axial and radial coordinate are normalized with the maximum domain lengths, $0.88$~m and $0.20$~m, respectively.}}\label{fig:ROM_JFM_line}
\end{figure*}

For this case, $N_t=3475$ training data points were sampled from the LES velocity fields, while $N_c=2500$ collocation points were selected by Latin Hypercube Sampling. The location of collocation and training points are shown as blue and yellow circles in Fig.~\ref{fig:colloc_ROM_JFM}, respectively. Spatial resolution was increased in the vicinity of the region immediately downstream of the nozzle exit ($\tilde{x}\approx 0$), since the jet is most pronounced there. The parameters of the PINN are then found by minimizing the composite loss function as explained in Section~\ref{sec:PINN}.

The first two diagrams in Fig.~\ref{fig:ROM_JFM} show the mean axial and mean radial velocity. The two plots show an excellent match between the LES and PINN fields. This is also evident in the more detailed examination of selected velocity profiles in Figs.~\ref{fig:ROM_JFM_lineA} \revise{and~\ref{fig:ROM_JFM_line_ur}}, where profiles of the radial and axial velocity components at \rev{three} axial positions are shown. It should be emphasized that the velocity fields were provided to the PINN in the form of training data. The excellent match between the velocity fields from LES and PINN, therefore, only shows that the minimization of the composite loss function in the PINN training has resulted in fields that match the training data. However, in addition to this, the PINN also identified the unknown shear stress term shown in Fig.~\ref{fig:ROM_JFMc}, for which no training data was provided to the PINN. 

The $\tau$-field is not resolved by the LES and has been identified by the PINN method solely based on the velocity data and the simplified equation of motion for circular jets. As it is not resolved by the LES, it can not be validated directly. However, the Reynolds stress term $\overline{u_x' u_r'}$, can be determined from the LES snapshots. As discussed in the above derivation, it is considered the most dominant shear stress term for circular jets. To validate the identified shear stress field, it can be compared to the shear stress from the LES. Fig.~\ref{fig:ROM_JFMc} shows the LES Reynolds stress term and the PINN identified $\tau$ field in the upper and lower half, respectively. Further details are provided in the form of radial profiles at \rev{three} axial positions in individual line plots in Fig.~\ref{fig:ROM_JFM_lineB}. The two fields show a very good agreement with small deviations in the shear layer region. 
At this point, we would like to reiterate that perfect agreement between the two fields is not to be expected, since the LES-Reynolds stress field is only an approximation to the shear stress term, as the equations used for identification are based on several assumptions. It is well known that the underlying assumptions are not particularly strong in the far field, and in these regions the agreement between LES and PINN is particularly good (see e.g. line plot at $\tilde{x}=0.5$ in Fig.~\ref{fig:ROM_JFM_lineB}). 

\begin{figure*}[t]
    \centering
    \includegraphics[width=\linewidth]{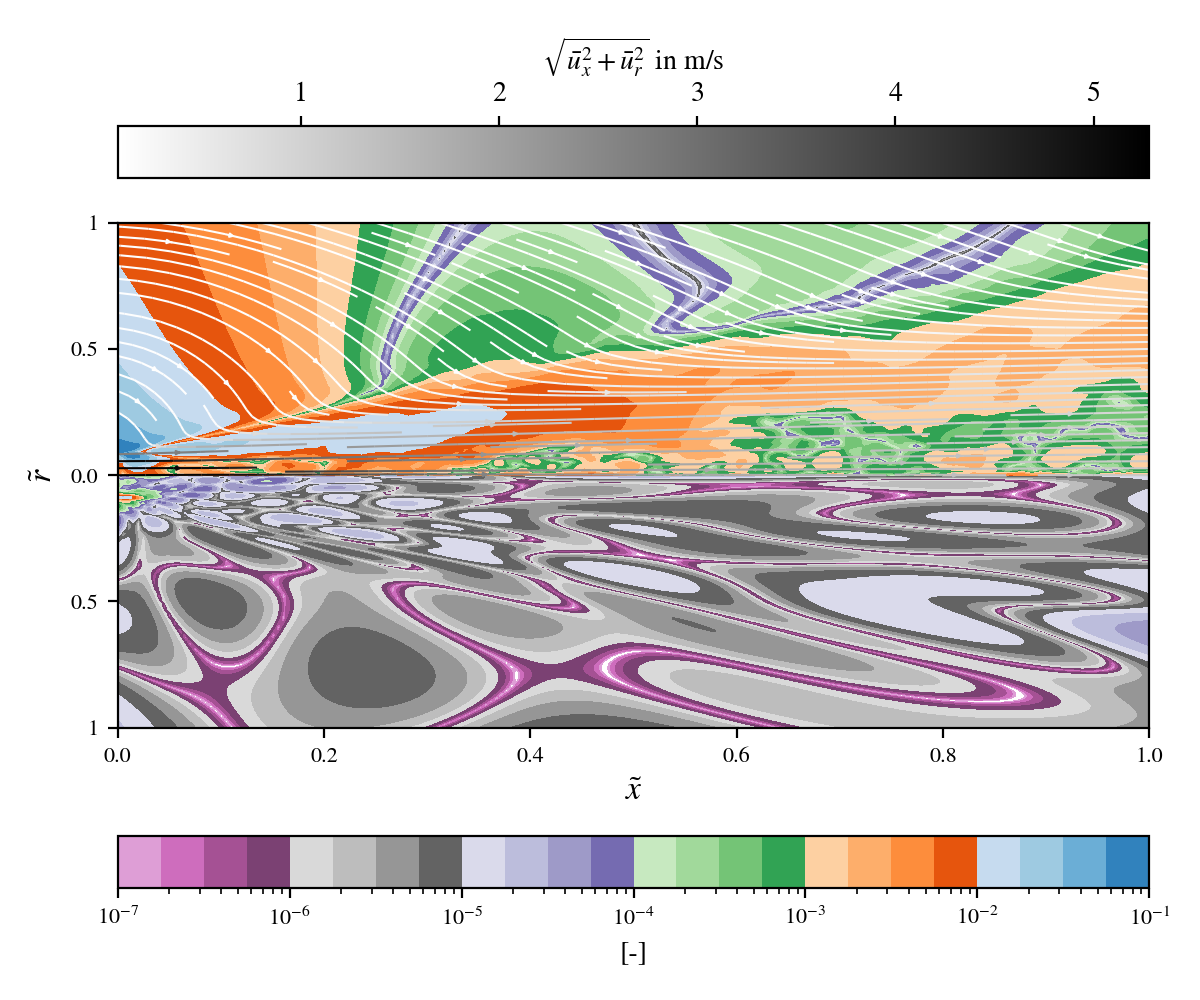}
    \caption{\rev{Shear stress term error analysis. Top: Normalized validation error $E_\tau$ and streamlines of the flow. Bottom: Normalized PDE residual $e_\tau$. The color bar at the top indicates the magnitude of the velocity along the streamlines. The color bar at the bottom shows the magnitude of the validation error and PDE residual.} \revise{The axial and radial coordinate are normalized with the maximum domain lengths, $0.88$~m and $0.20$~m, respectively.}}\label{fig:error_POF3rev}
\end{figure*}

\rev{For a more detailed analysis of the result, we show the normalized residual of the PDE $e_\tau \coloneqq |r \mathcal{N}_\tau/U_x^2|$ and the normalized validation error $E_\tau \coloneqq |\tau^\alpha + \overline{u_x'u_r'}\revise{/T}|$ on an equidistant grid with 40000 points in the lower and upper half of Fig.~\ref{fig:error_POF3rev}, respectively; as a reminder, at 2500 collocation points information about the PDE was provided to the PINN during training. The upper half of the figure also shows the streamlines of the flow. It can be observed that the axial velocity does not drop to zero at the left wall, which is due to the fact that a purge flow was used in the LES to avoid numerical problems. However, as can be seen from the color bar of the streamlines, the absolute value of the velocity in this region is very low.}

\rev{The residual of the PDE, shown in the lower half of Fig.~\ref{fig:error_POF3rev}, is very low in most areas; the spatial mean is $9.5\times10^{-6}$. A peak can be observed at the point of maximum gradient of the velocity field, right at the nozzle exit. Here, the maximum value of $1.8\%$ is reached. When the residual is evaluated at the collocation points only (used for training), the maximum value drops by two orders of magnitude, indicating that not enough collocation points were chosen in this region to adequately approximate the gradients.}

\rev{The normalized validation error $E_\tau$, shown in the upper half of Fig.~\ref{fig:error_POF3rev}, is $3.1\times10^{-3}$ in the spatial mean. Two distinct areas can be observed where the deviation is greater than $1\%$: in the left part of the domain, in a region near the wall and further downstream of the nozzle in the shear layer. The maximum deviation is $9.4\%$ and is located near the nozzle exit where the two areas merge. Unlike the first case, where the data (velocities) were compatible with the PDE (continuity equation), the discrepancies in this case are not only due to the assimilation methodology, but may also be due to the fact that the PDE is a simplified model (boundary layer equation) that is not necessarily compatible with the data in all regions. Both regions showing an increased deviation are in areas where the assumption of neglecting the axial gradient of the radial velocity component is not well satisfied; moreover, $\bar{u}_x \gg \bar{u}_r$ is not satisfied in the outer region near the inlet, see Fig.~\ref{fig:ROM_JFMa} and \ref{fig:ROM_JFMb} and streamlines in Fig.~\ref{fig:error_POF3rev}. Further downstream in the far field, a good approximation to the Reynolds stress term is observed. As the streamlines clearly show, a parallel flow field is approximated in this region, which satisfies the assumption of the boundary layer equations. Thus, it can be presumed that the observed deviations in the upstream region are largely due to the simplified model applied.}

The overall good qualitative and quantitative agreement leads us to conclude that the PINN method is capable of assimilating mean fields using simplified models and high-fidelity mean fields, including the identification of closure fields. \rev{For linearized mean field methods, which often lack suitable closure fields, this is a key finding.}

\subsection{The swirling jet - mean flow assimilation based on RANS equations}
\rev{Having successfully shown that PINNs can assimilate hidden quantities based on reduced sets of equations that are not necessarily satisfied by the data, we further increase the complexity in this final example.} Therein, we consider experimental velocity data from PIV measurements of a strongly non-parallel swirling jet and assimilate multiple mean flow quantities based on RANS equations. \rev{This example resembles a typical problem that often arises when linearized mean field methods are to be applied, but the measured data under consideration are limited to certain quantities and the turbulent closure field is missing.}

Swirling jets are an essential component of many technical products; since they lead to higher levels of turbulence compared to non-swirling jets, they are used, for example, to improve mixing processes.~\cite{Liang2005} If the swirl is sufficiently strong, an inner recirculation zone, also called a vortex breakdown bubble, is formed shortly after the nozzle exit. In stationary gas turbines, this region of low axial momentum is often used to stabilize the flame in the flow.~\cite{Huang2009,Reumschussel2022} In contrast to the round jet considered so far, the swirl flow features an azimuthal velocity component and is strongly non-parallel. As a consequence, the momentum transport in all spatial directions has to be considered to model the mean flow. In addition, three-dimensional shear occurs, which in turn leads to strong anisotropic turbulence, i.e., \rev{all components of the Reynolds stress tensor are relevant, in contrast to weakly non-parallel flows with only one relevant component.}

PIV measurements are commonly used to study the flow characteristics of jets. However, this technique is particularly complex for the acquisition of swirling configurations, since the flow fields have a non-negligible azimuthal velocity component. The acquisition of the three-dimensional flow is only possible with considerable additional effort. One way to obtain mean fields of all $\bar{\mathbf{u}}$ components is to examine individual sections of the flow field one after the other. Another possibility requiring two cameras is the stereo PIV method which can resolve all three velocity components in a two-dimensional laser plane, yielding the complete mean flow, given axis-symmetry. Both options involve considerable additional effort compared to the acquisition of flows without an azimuthal velocity component. \rev{Thus, the possibility of using PINNs to assimilate the azimuthal velocity component from the axial and radial components is explored in the following.} 

In this last example, we consider PIV data of an axis-symmetric swirling jet acquired using stereo PIV measurements. The data were measured on a swirling jet at a Reynolds number of 4000 and a swirl number of 0.98. Further details on the experimental setup can be found in the original publication of the data.~\cite{Sieber.2021} Using only the 2D mean flow field, consisting of the axial and radial velocity component and limited information about the azimuthal velocity component, it is investigated whether the complete mean flow field can be assimilated using the PINN methodology. For this purpose, the axis-symmetric RANS equations are considered. The commonly applied Boussinesq hypothesis is used to relate the Reynolds stresses to the mean flow gradient. For highly three dimensional flows like the swirling jet, the Boussinesq hypothesis is a strong assumption, which makes the RANS equations based on the Boussinesq turbulence model a simplified model. To use the RANS equations, the pressure and the eddy viscosity field are assimilated simultaneously, in addition to the azimuthal velocity component. Hence, in this case, the DNN maps the \revise{normalized} spatial coordinates to all \revise{normalized} velocity components, pressure, and eddy viscosity
\begin{align}
    \left[\bar{u}_x^\alpha(\tilde{x},\tilde{r}),\,\bar{u}_r^\alpha(\tilde{x},\tilde{r}),\,\bar{u}_\theta^\alpha(\tilde{x},\tilde{r}),\,\bar{p}^\alpha(\tilde{x},\tilde{r}),\,\nu_T^\alpha(\tilde{x},\tilde{r}) \right]^\text{T} = \mathbf{f}_{\alpha}(\tilde{x},\tilde{r}).
\end{align}
Compared to the previous examples, the complexity has increased considerably. First, we consider real measurement data, and second, we assimilate three unknown fields simultaneously using RANS equations as a simplified model. 

For an incompressible flow, the RANS equations read
\begin{align}
     (\bar{\mathbf{u}} \cdot \nabla) \bar{\mathbf{u}} + \frac{1}{\rho} \nabla \bar{p} -  \nabla \cdot (\nu_T (\nabla \bar{\mathbf{u}} + (\nabla \bar{\mathbf{u}})^\text{T})) &= \mathbf{0} = \begin{bmatrix}
        \mathcal{N}_x(\bar{u}_x,\bar{u}_r,\bar{p},\nu_T) \\
        \mathcal{N}_r(\bar{u}_x,\bar{u}_r,\bar{u}_\theta,\bar{p},\nu_T) \\
        \mathcal{N}_\theta(\bar{u}_r,\bar{u}_\theta,\nu_T)
     \end{bmatrix} \label{eq:swirl_MOM}\\
      \nabla \cdot \bar{\mathbf{u}} &= 0 = \mathcal{N}_c(\bar{u}_x, \bar{u}_r) \label{eq:swirl_MASS}
\end{align}
where $\bar{\mathbf{u}}=[\bar{u}_x\,\bar{u}_r\,\bar{u}_\theta]^\text{T}$ is the mean velocity vector that contains the axial, radial and azimuthal velocity component and the Boussinesq hypothesis has been applied to relate the Reynolds stresses to the mean flow gradient. Note that the mean flow is axis-symmetric, such that all spatial derivatives with respect to the azimuthal coordinate are zero. The individual components of the momentum equations ($\mathcal{N}_x, \, \mathcal{N}_r, \, \mathcal{N}_\theta$)\rev{, which provide information about the coupling of the velocity components,} can be found in Appendix A.

\begin{figure*}[b]
    \centering
    \includegraphics[width=\linewidth]{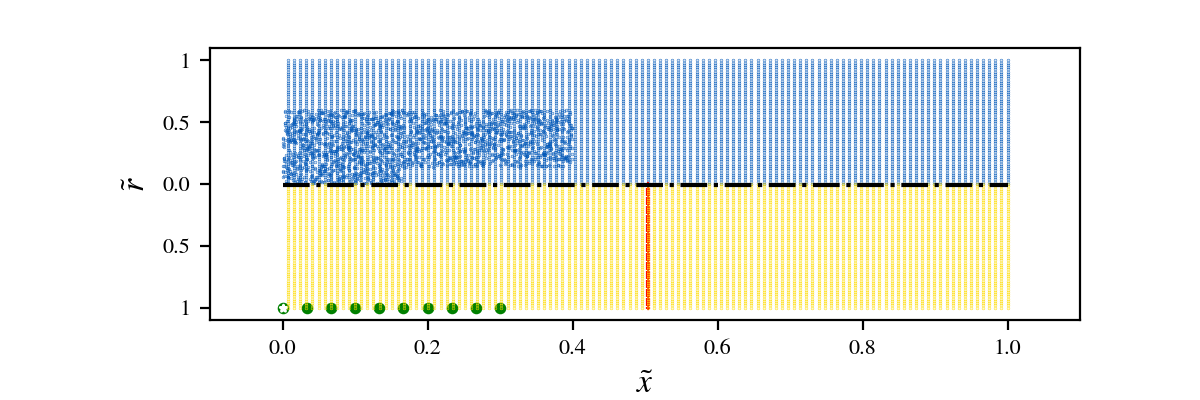}
    \caption{\rev{Computational domain of the swirling jet.} Yellow: Locations at which radial and axial velocity data is provided for PINN training. Red: Training points for radial velocity data. Green: Locations at which $\bar{u}_\theta = 0$ is provided for training. White star: location at which $\bar{p} = 0$ is set. Blue: Collocation points, at which the fulfillment of the PDEs is checked. \revise{The axial and radial coordinate are normalized with the maximum domain lengths, $0.20$~m and $0.09$~m, respectively.}}\label{fig:colloc_swirl_JFM}
\end{figure*} 
The RANS equations, Eqs.~\ref{eq:swirl_MOM}-\ref{eq:swirl_MASS} consist of four equations and five unknowns, namely the three velocity components, the pressure and the eddy viscosity. Hence, a closure model relating the eddy viscosity to the velocity components is usually required to solve the set of equations. In the given problem, however, the azimuthal velocity component, the pressure field and the eddy viscosity are to be assimilated from given radial and axial velocity fields, reducing the number of unknowns to three. Accordingly, the continuity equation could be omitted. However, to ensure that the PINN approximations of the mean flow of $\bar{u}_x$ and $\bar{u}_r$ also satisfy continuity, it is included in the set of equations here. 

\begin{figure*}[!ht]
    \centering
    \begin{subfigure}{\linewidth}
        \includegraphics[width=0.8\linewidth]{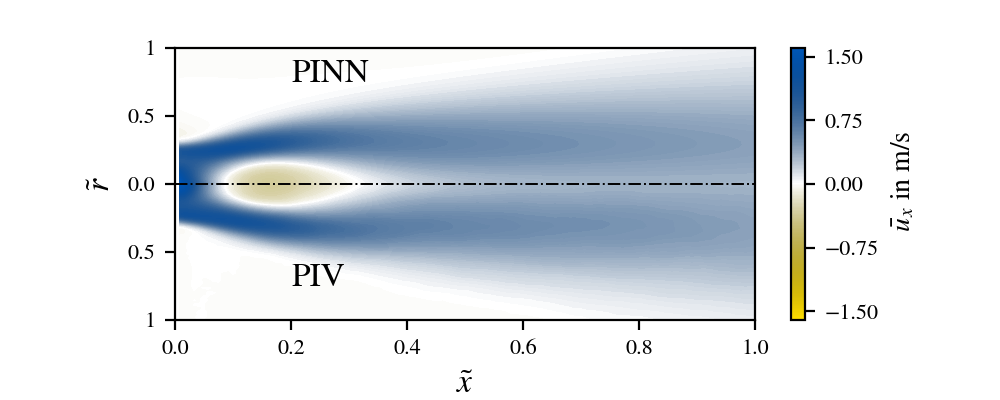}
        \caption{\revise{PINN and LES axial velocity}}\label{fig:swirl_JFMa}
    \end{subfigure}
    \begin{subfigure}{\linewidth}
        \includegraphics[width=0.8\linewidth]{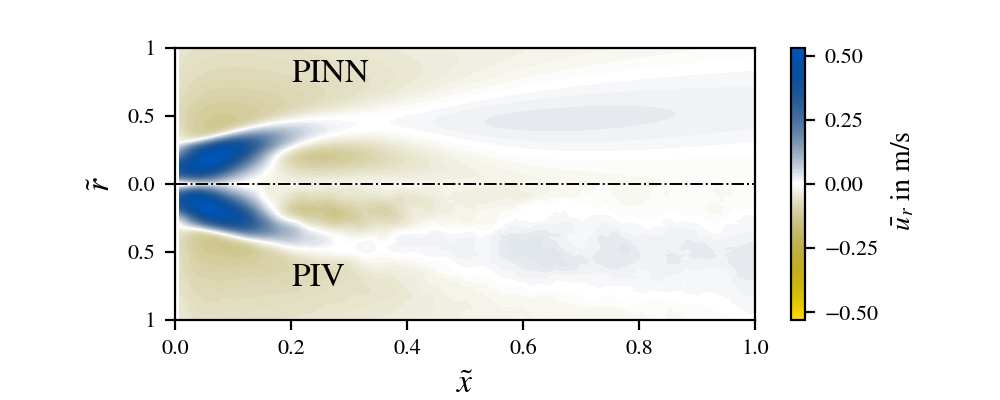}
        \caption{\revise{PINN and LES radial velocity}}\label{fig:swirl_JFMb}
    \end{subfigure}
    \begin{subfigure}{\linewidth}
        \includegraphics[width=0.8\linewidth]{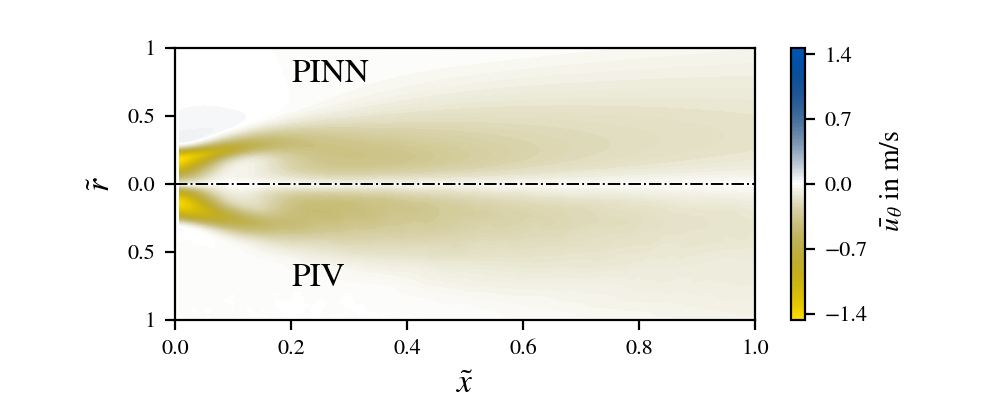}
        \caption{\revise{PINN and LES azimuthal velocity}}\label{fig:swirl_JFMc}
    \end{subfigure}
    \caption{\revise{Mean velocity fields.} (a) axial, (b) radial and (c) azimuthal velocity. Upper halves show the PINN results, lower halves show the respective PIV validation data. \revise{The axial and radial coordinate are normalized with the maximum domain lengths, $0.20$~m and $0.09$~m, respectively.}}\label{fig:swirl_JFM}
\end{figure*}

\rev{Fig.}~\ref{fig:colloc_swirl_JFM} shows the location of collocation and training points used for PINN training in the upper and lower part, respectively. At $N_t=6664$ points, the radial and axial mean velocity components are used as training data, shown in yellow in Fig.~\ref{fig:colloc_swirl_JFM}. The corresponding PIV mean velocity fields are shown in the lower halfs of \rev{Fig.~\ref{fig:swirl_JFMa} and Fig.~\ref{fig:swirl_JFMb}}. The vortex breakdown and central recirculation zone are evident from the velocity fields. The lower half of \rev{Fig.~\ref{fig:swirl_JFMc}} shows the measured azimuthal velocity component which is used for validation only. Collocation points are placed at the locations of the training points and at 2000 additional positions near the nozzle exit, in regions of high absolute axial and radial velocity. The additional points are selected by means of Latin Hypercube Sampling. At these total $N_c=8664$ collocation points, shown in blue in Fig.~\ref{fig:colloc_swirl_JFM}, the PINN is evaluated for fulfillment of the PDEs, Eqs.~\ref{eq:swirl_MOM}-\ref{eq:swirl_MASS}. 

\begin{figure*}[hb]
    \centering
    \begin{subfigure}{0.22\linewidth}
        \includegraphics[width=\linewidth]{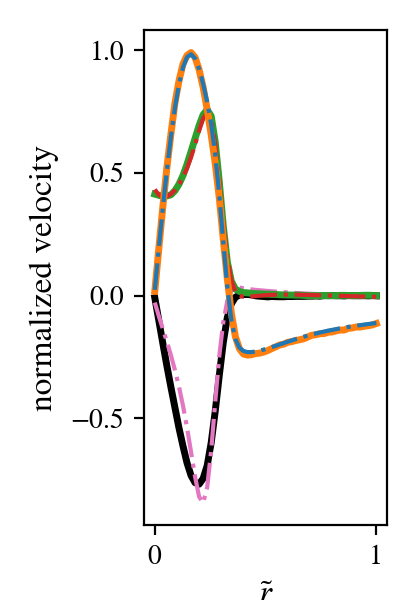}
        \caption{$\tilde{x}=0.05$}
    \end{subfigure}
    \begin{subfigure}{0.21\linewidth}
        \includegraphics[width=\linewidth]{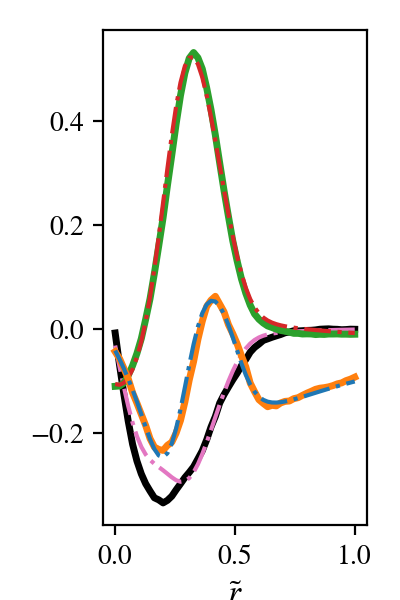}
        \caption{$\tilde{x}=0.25$}
    \end{subfigure}
    \begin{subfigure}{0.21\linewidth}
        \includegraphics[width=\linewidth]{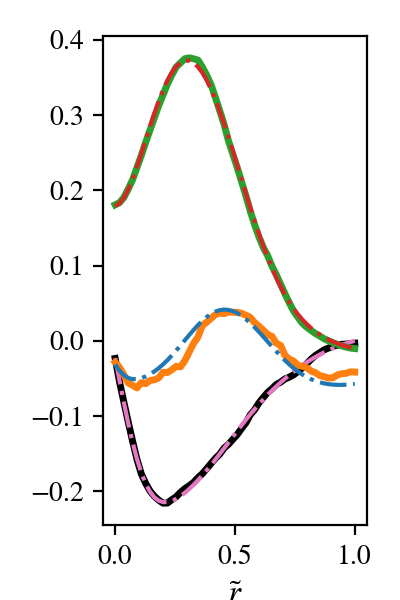}
        \caption{$\tilde{x}=0.5$}\label{fig:swirl_line_JFMc}
    \end{subfigure}
    \begin{subfigure}{0.315\linewidth}
        \raisebox{-4.5cm}{\includegraphics[width=\linewidth]{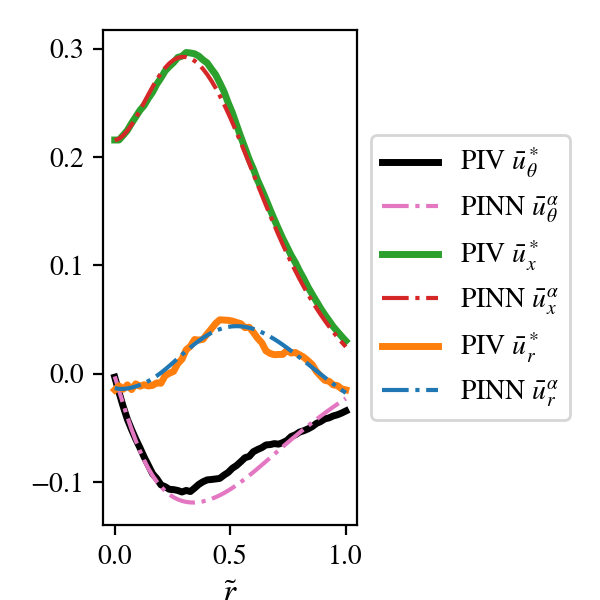}}
        \caption{$\tilde{x}=0.9$}
    \end{subfigure}
    \caption{Normalized axial, radial and azimuthal velocity profiles at four axial positions; (a) $\tilde{x}=0.05$, (b) $\tilde{x}=0.25$, (c) $\tilde{x}=0.5$, (d) $\tilde{x}=0.9$. Dashed and solid lines show PINN and PIV quantities, respectively. \revise{The axial and radial coordinate are normalized with the maximum domain lengths, $0.20$~m and $0.09$~m, respectively. The axial, radial and azimuthal velocity are normalized with $U_x=1.57$~m/s, $U_r=0.53$~m/s and $U_\theta = 1.42$~m/s, respectively.}}\label{fig:swirl_line_JFM}
\end{figure*}

Identification of the mean flow based on the radial and axial components alone has proven difficult (not shown). In regions where the azimuthal velocity is large, i.e., near the nozzle exit and around the central recirculation zone, the identification works \rev{qualitatively} well, but in the far field the identified azimuthal velocity is significantly overestimated by the PINN. We suspect that the coupling between the equations is not strong enough in the far field and that the training data without azimuthal component contain insufficient information about turbulent dissipation due to the azimuthal component ($\overline{u_r'u_\theta'}$ and $\overline{u_x'u_\theta'}$). To improve PINN assimilation, a radial velocity profile of the azimuthal component at $\tilde{x}=0.5$ is added to the training data. The $N_\theta=56$ points used for this purpose are shown in red in Fig.~\ref{fig:colloc_swirl_JFM}. The corresponding velocity profile is shown in \rev{Fig.~\ref{fig:swirl_line_JFMc}}. The azimuthal velocity profile at $\tilde{x}=0.5$ drops to 0 at $\tilde{r}=1$. Therefore, it is a natural assumption that for all $\tilde{x}<0.5$ the azimuthal component is zero at $\tilde{r}=1$. In order to incorporate further information into the training, the boundary condition $\bar{u}_\theta = 0$ at $\tilde{r}=1$ is imposed at $N_{bc}=10$ points, which are shown in green in Fig.~\ref{fig:colloc_swirl_JFM}. 
Since the PDEs contain only a pressure gradient, the absolute pressure level cannot be estimated based on the RANS equations. To set the level to a unique value, we set a boundary condition for the pressure at one point $\bar{p}(0,1)=0$, shown as a white star in Fig.~\ref{fig:colloc_swirl_JFM}. This results in the following two loss terms, which are incorporated into the composite loss function,
\begin{align}
\begin{split}
     ||L_{\text{train}}|| &=  \frac{1}{2N_t} \sum_{i=1}^{N_t}(\bar{u}_x^*(\tilde{\mathbf{x}}_i) - \bar{u}_x^\alpha(\tilde{\mathbf{x}}_i))^2+(\bar{u}_r^*(\tilde{\mathbf{x}}_i) - \bar{u}_r^\alpha(\tilde{\mathbf{x}}_i))^2 + ...\\
     &\quad \frac{1}{N_\theta} \sum_{i=1}^{N_\theta}(\bar{u}_\theta^*(0.5,r_i) - \bar{u}_\theta^\alpha(0.5,r_i))^2 + ... \\
     &\quad \frac{1}{N_{bc}} \sum_{i=1}^{N_{bc}}(\bar{u}_\theta^*(x_i,1) - \bar{u}_\theta^\alpha(x_i,1))^2+ ...\\
     &\quad +(\bar{p}^\alpha(0,1))^2, \end{split}\\
\begin{split}
     ||L_{\text{PDE}}|| &= \frac{1}{N_c}\sum_{i=1}^{N_c} (r\mathcal{N}_x(U_x\bar{u}_x^\alpha(\tilde{\mathbf{x}}_i),\,U_r\bar{u}_r^\alpha(\tilde{\mathbf{x}}_i),\,P\bar{p}^\alpha(\tilde{\mathbf{x}}_i),\,N_U\nu_T^\alpha(\tilde{\mathbf{x}}_i))^2 + ... \\
     &\quad \frac{1}{N_c}\sum_{i=1}^{N_c} (r\mathcal{N}_r(U_x\bar{u}_x^\alpha(\tilde{\mathbf{x}}_i),\,U_r\bar{u}_r^\alpha(\tilde{\mathbf{x}}_i),\,U_\theta\bar{u}_\theta^\alpha((\tilde{\mathbf{x}}_i)),\,P\bar{p}^\alpha(\tilde{\mathbf{x}}_i),\,N_U\nu_T^\alpha(\tilde{\mathbf{x}}_i))^2 + ... \\
     &\quad \frac{1}{N_c}\sum_{i=1}^{N_c} (r\mathcal{N}_\theta(U_r\bar{u}_r^\alpha(\tilde{\mathbf{x}}_i),\,U_\theta\bar{u}_\theta^\alpha((\tilde{\mathbf{x}}_i)),\,N_U\nu_T^\alpha(\tilde{\mathbf{x}}_i))^2 + ... \\
     &\quad \frac{1}{N_c}\sum_{i=1}^{N_c} (r\mathcal{N}_c(U_x\bar{u}_x^\alpha(\tilde{\mathbf{x}}_i),\,U_r\bar{u}_r^\alpha(\tilde{\mathbf{x}}_i))^2,
     \end{split}
\end{align}
where the PDEs are multiplied with the radial coordinate again and are formulated in physical units. \rev{$\alpha$ and $*$ superscripts denote PINN and PIV mean flow quantities, respectively.} \revise{For the normalization of axial and radial velocity we use the respective maximum measured values, $U_x=1.57$~m/s and $U_r=0.53$~m/s .} The maximum values for the azimuthal velocity component, eddy viscosity and pressure ($U_\theta, \, P, \, N_U$) are unknown and must be estimated. Here, we use the maximum measured azimuthal velocity \revise{$U_\theta=1.42$~m/s} and estimate $P$ and $N_U$ to be 20 kg/ms$^2$ and \revise{$10^{-5}$} m$^2$/s, respectively. PINN training is performed as described in Section~\ref{sec:PINN}. Compared to the previous examples, significantly more derivatives have to be computed. Due to the increased computational cost, only $15000$ L-BFGS iterations are performed in this case.

\begin{figure*}[ht]
    \centering
    \begin{subfigure}{1\linewidth}
        \includegraphics[width=0.75\linewidth]{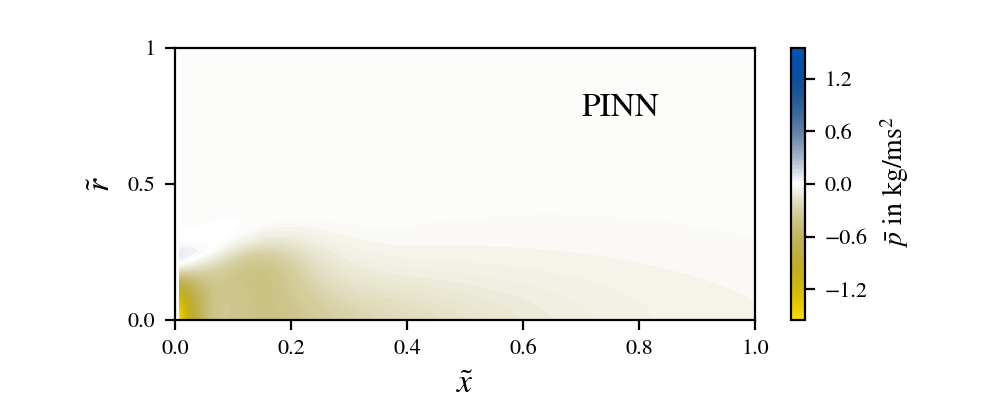}
        \caption{\revise{PINN identified} pressure}
    \end{subfigure}
    \begin{subfigure}{1\linewidth}
        \includegraphics[width=0.75\linewidth]{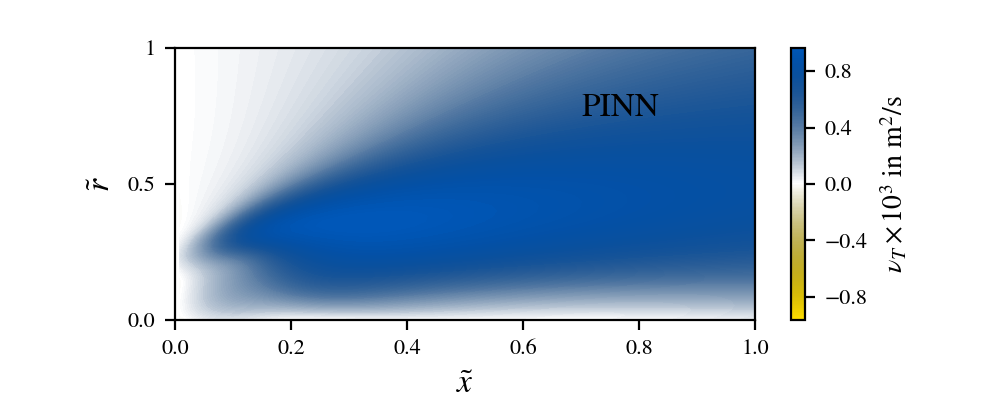}
        \caption{\revise{PINN identified} eddy viscosity}
    \end{subfigure}
    \caption{\revise{PINN identified mean fields.} (a) pressure and (b) eddy viscosity. \revise{The axial and radial coordinate are normalized with the maximum domain lengths, $0.20$~m and $0.09$~m, respectively.}}\label{fig:swirl_JFM2}
\end{figure*} 

The lower halves of the subplots in Fig.~\ref{fig:swirl_JFM} show the velocity fields assimilated using the PINN method. \rev{The subplots in Fig.~\ref{fig:swirl_JFM2} show the assimilated (a) mean pressure and (b) eddy viscosity field.} Radial velocity profiles at four axial positions are also shown for all three velocity components in Fig~\ref{fig:swirl_line_JFM}. The PINN method succeeds in assimilating the velocity fields using the training data and the RANS equations with high accuracy. Minor discrepancies between measured data and the PINN output values can be observed especially for the azimuthal component. \rev{A close inspection of the line plots shows that the PINN assimilates noise-free velocity fields, even though the measured velocity fields used for training are noisy. This results mostly from the imposed conformity with the physical equations, which leads to the assimilation of differentiable fields.}

\begin{figure*}[b]
    \centering
    \includegraphics[width=\linewidth]{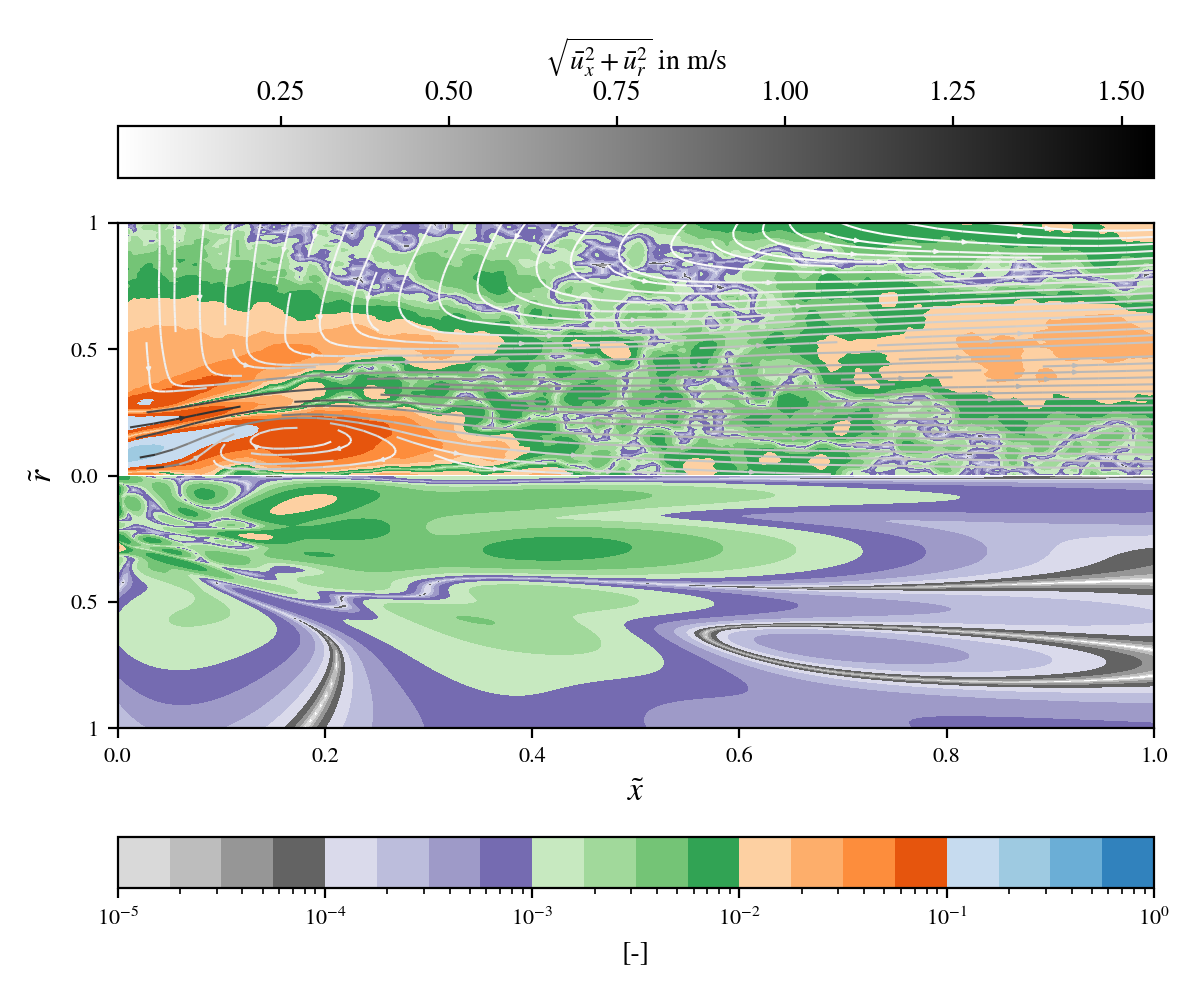}
    \caption{\rev{Azimuthal velocity component error analysis. Top: Normalized validation error $E_\theta$ and streamlines of the flow. Bottom: Normalized PDE residual $e_\theta$. The color bar at the top indicates the magnitude of the velocity along the streamlines. The color bar at the bottom shows the magnitude of the validation error and PDE residual. \revise{The axial and radial coordinate are normalized with the maximum domain lengths, $0.20$~m and $0.09$~m, respectively.}}}\label{fig:error_swirl}
\end{figure*}

\rev{For a more detailed analysis, we focus below on the azimuthal velocity component, for which little direct information was available for training and which was therefore identified predominantly from the PDEs. Fig.~\ref{fig:error_swirl} shows the normalized validation error $E_\theta \coloneqq |\bar{u}_\theta^\alpha - \bar{u}_\theta^*|$ \revise{(both quantities are normalized)} and the streamlines of the flow in the upper half. The lower half shows the normalized residual of the azimuthal momentum equation $e_\theta \coloneqq |r\mathcal{N}_\theta / U_\theta^2|$. The azimuthal momentum equation was chosen as an example; the corresponding distributions of residuals of the other PDEs are very similar. Both quantities were evaluated at 40000 grid points on an equidistant grid.}

\rev{The residual of the PDE is in orders of magnitude between $10^{-5}$ and $10^{-2}$ (spatial mean is $1.4\times 10^{-3}$) and shows no noticeable regions with high values; the maximum value is $2.1\%$. The validation error, on the other hand, has a mean value of $1.2\%$ and shows clearly elevated values at the inlet, which decrease along the flow but are still visible in the outer shear layer and in the recirculation zone throughout the front third of the domain. The maximum deviation at the inlet is $21\%$. The reasons for the high deviation in the upstream region may be manifold. Since we cannot identify any conspicuous regions in the PDE residuals, we assume that the methodology has generally worked well and that changing the hyperparameters, e.g., considering more collocation points, would not reduce the deviation. It can be seen from Fig.~\ref{fig:error_swirl} that in regions where information about the azimuthal velocity component is available during training ($\hat{x}=0.5 $, $\hat{r}=1$), the validation error is also low. Thus, it stands to reason that the discrepancy in the inlet region can be reduced if the PINN is provided with more information about the component in this region. However, even then there is no guarantee that a solution can be found that also satisfies the RANS equations. In the upstream regions where the deviation is high, the degree of anisotropic turbulence is also high. It is therefore likely that with additional information about the azimuthal component the validation error decreases, but at the same time the Boussinesq model reaches its limits and the PDE residuals increase accordingly.}

At this point, we want to remark that the measurement data are also subject to errors and do not necessarily represent the physical truth. \rev{The measurement noise, which is evident in Fig.~\ref{fig:error_swirl} from the jagged edges of the surface levels, is a major contributor to the mean validation error being higher than in the previously considered cases.} As discussed above, the PINN assimilation method does involve intrinsic denoising, which should be kept in mind when considering the results. The fact that the azimuthal velocity field can be assimilated with \rev{very high precision in most regions of the domain}, based on the given training data is an exceptional result with direct implications for future experimental analysis of axis-symmetric flows. Since only one azimuthal profile was used, two measurement planes will suffice in the future to capture the entire velocity field: one along the jet axis to measure the axial and radial component, and one measurement plane perpendicular to the jet axis to measure one azimuthal profile. The remaining azimuthal component can then be assimilated as shown here. Compared to conventional methods, this approach can significantly save measurement effort while providing a deeper insight into the physics of the flow. 

No measured data are available for the eddy viscosity and the mean pressure field, meaning that they cannot be directly validated. The pressure field looks very similar to what is typically observed for swirling jets,~\cite{percin2017} with a region of low pressure directly behind the nozzle exit in the area of recirculation bubble. 

The eddy viscosity relates the mean flow gradient to the Reynolds stresses via the Boussinesq hypothesis. For an axis-symmetric flow, the corresponding equations are
\begin{align}
    \overline{u_x'u_r'} &= - \nu_T \left(\frac{\partial \bar{u}_x}{\partial r}+\frac{\partial \bar{u}_r}{\partial x}\right) \label{eq:B1} \\
    \overline{u_x'u_\theta'} &= - \nu_T \frac{\partial \bar{u}_\theta}{\partial x} \\
    \overline{u_r'u_\theta'} &= - \nu_T \left(\frac{\partial \bar{u}_\theta}{\partial r}-\frac{\bar{u}_\theta}{r}\right)\label{eq:B3} 
\end{align}
Since the Reynolds stresses can be determined from the PIV snapshots, they can be used to validate the eddy viscosity determined with the PINN. By exploiting the fact that the PINN is differentiable, the modeled Reynolds stresses can be calculated directly using Eqs.~\ref{eq:B1}-\ref{eq:B3}.

\begin{figure*}[t!]
    \centering
    \begin{subfigure}{\linewidth}
        \includegraphics[width=0.8\linewidth]{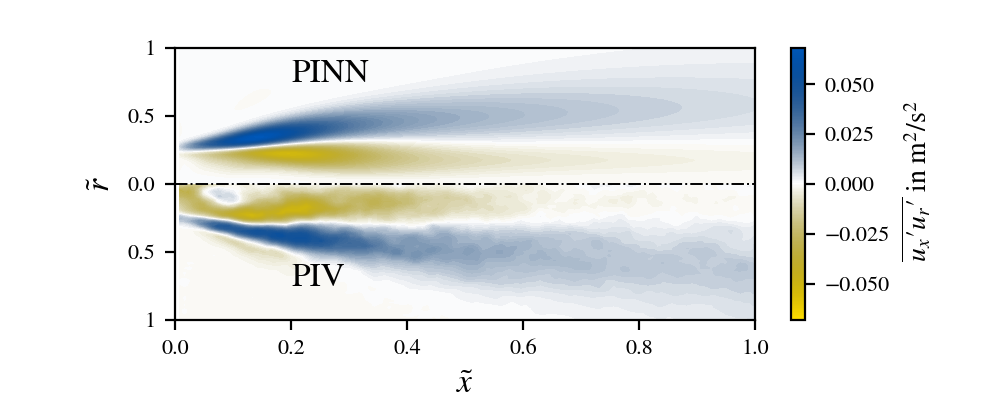}
        \caption{axial-radial component}\label{fig:swirl_JFM_Reynoldsa}
    \end{subfigure}
    \begin{subfigure}{\linewidth}
        \includegraphics[width=0.8\linewidth]{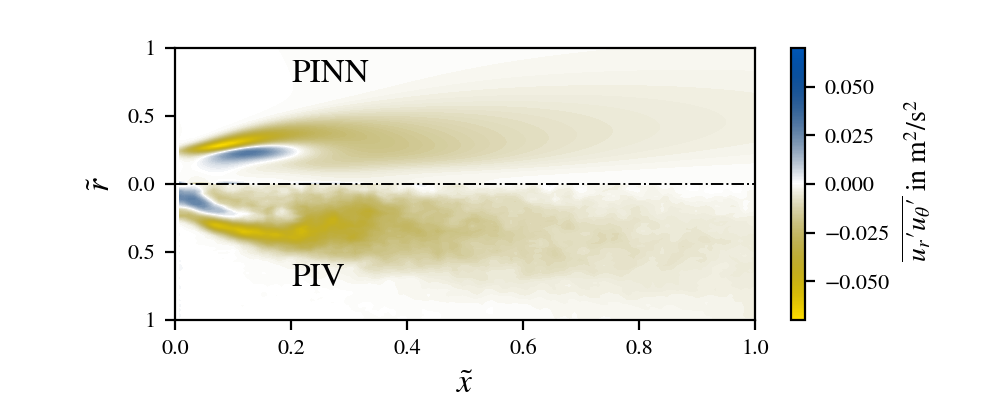}
        \caption{radial-azimuthal component}\label{fig:swirl_JFM_Reynoldsb}
    \end{subfigure}
    \begin{subfigure}{\linewidth}
        \includegraphics[width=0.8\linewidth]{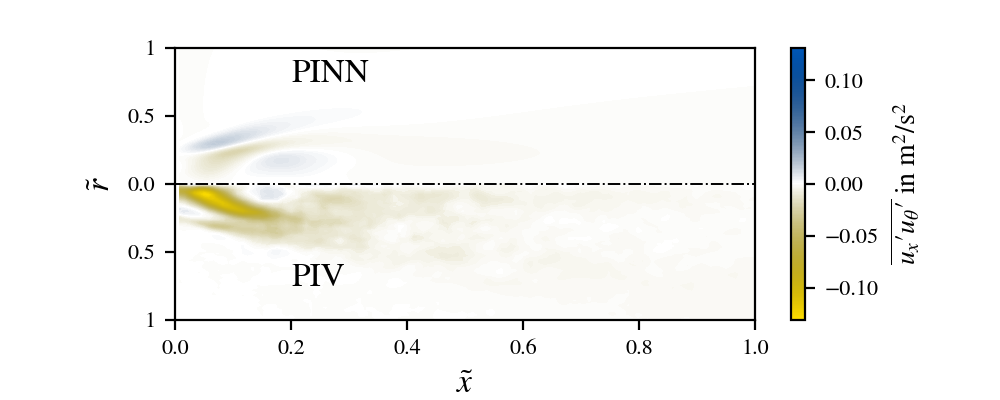}
        \caption{axial-azimuthal component}\label{fig:swirl_JFM_Reynoldsc}
    \end{subfigure}
    \caption{\rev{Reynolds stress components. (a) axial-radial $\overline{u_x'u_r'}$, (b) radial-azimuthal $\overline{u_r'u_\theta'}$ and (c) axial-azimuthal $\overline{u_x'u_\theta'}$ component. Upper halves show the PINN results, lower halves show the measured data.} \revise{The axial and radial coordinate are normalized with the maximum domain lengths, $0.20$~m and $0.09$~m, respectively.}}\label{fig:swirl_JFM_Reynolds}
\end{figure*}

\rev{Fig.}~\ref{fig:swirl_JFM_Reynolds} shows the measured and modeled Reynolds stresses, $\overline{u_x'u_r'}$, $\overline{u_r'u_\theta'}$ and $\overline{u_x'u_\theta'}$ from top to bottom. The respective upper halves show the components identified on the Boussinesq model and the PINN, the lower halves show the corresponding measured components. The axial-radial component, \rev{Fig.~\ref{fig:swirl_JFM_Reynoldsa}}, can be well approximated by the PINN. In the radial-azimuthal component, \rev{Fig.~\ref{fig:swirl_JFM_Reynoldsb}}, the mismatch is more pronounced, but the Reynolds stress is still qualitatively well represented by the PINN. The modeled axial-azimuthal component however, \rev{Fig.~\ref{fig:swirl_JFM_Reynoldsc}}, deviates significantly from the corresponding measured value. 

The fact that the axial-radial component is best approximated is probably due to the training data, which includes information on the mean axial and radial velocity components throughout the entire domain. This allows the turbulent dissipation in the axial-radial direction to be appropriately estimated. 
Information about the radial-azimuthal dissipation is included in the training data merely through the given radial $\bar{u}_\theta$ profile, which seems to be sufficient for an approximate modeling.
In contrast, however, no data at all was provided for the axial-azimuthal component (there is no information on the axial gradient of the azimuthal component in the training data) and thus a rather deficient approximation of this component is not surprising. However, it should also be noted that a difference in Reynolds stresses between PINN and PIV may occur for two reasons. On the one hand, the PINN may not identify the corresponding component well; on the other hand, the underlying assumptions of the Boussinesq model may not be satisfied in certain regions, such that the PINN cannot estimate the Reynolds stresses well in these regions due to the model structure. For the investigated swirling jet, for example, it was shown that the experimental setup is prone to low frequency perturbations of the incoming flow that cause an axial shift of the upstream end of the recirculation zone.\cite{Rukes.2015,Lang.2017} This leads to large values of Reynolds stresses at this point, but these only represent slow changes in the flow due to changing inflow conditions. Consequently, these Reynolds stresses do not contribute to turbulent diffusion and are therefore not identified as contributions in the Boussinesq model. 

This case shows that PINNs are very well suited to assimilate mean flows based on the RANS equations and measured fields. The application of PINNs to the RANS equations is of particular importance since they are the governing system of equations for time-averaged flow fields. \rev{The example presented illustrates how all required quantities for a linearized mean field analysis, including the closure field, can be assimilated from a limited data set, paving the way for the application of linearized methods to measured data. Moreover, the} identification of missing quantities can save measurement effort and further contribute to the modeling of closure terms.

\section{Conclusion}
In this study, we apply the PINN method for \rev{turbulent} mean flow assimilation using reduced sets of equations and simplified models. The former are understood as equations representing only parts of the underlying physics, while the latter are defined as approximate equations that are not necessarily satisfied by the high-fidelity training data. \rev{The objective of the study is to prepare and augment mean fields in such a way that they can be used for linearized mean field analysis. For this purpose, examples of increasing complexity are considered, showing typical applications where linear methods are often limited due to missing quantities.}

By use of reduced sets of equations we show the simplicity with which PINNs are able to identify hidden variables from basic phyical laws. 
Furthermore we demonstrate PINN-based data assimilation from simplified models and discuss requirements and limitations of the method.
The use of simplified models is of particular importance for the assimilation of \rev{turbulent} mean flows, since the RANS equations that are often used in fluid mechanics involve closure models that are typically based on strong assumptions and therefore must be considered a simplified model in this sense. This study shows how PINNs allow to augment measurement data with multiple field quantities, which can save considerable experimental costs. \rev{Moreover, we find that PINNs introduce an intrinsic denoising behavior. The} method can be exploited to identify closure terms, which will be very helpful \rev{for mean field analysis based on linearized methods and the future development of closure models}. Mean flow data assimilation based on PINNs has further significant advantages over conventional data assimilation methods. PINNs are easy to implement and can be quickly adapted to a wide range of applications. The methodology \rev{does not require a grid or discretization scheme} and the mean flow is assimilated in the form of a continuous and differentiable function - the PINN. 
From the results presented, the method is found to be very robust; the three considered applications differ significantly, but could all be solved successfully with the same network architecture, the same hyperparameters and the same optimization routine. In summary, assimilation of mean flow data with PINNs is a highly effective and exceedingly promising technique with the potential to significantly contribute to mean field-based research in fluid mechanics in the future. 

\section*{Appendix A}\label{sec:appendix}
\renewcommand{\theequation}{A.\arabic{equation}}
\setcounter{equation}{0}
\begin{align}
    0 =& \underbrace{\bar{u}_r \frac{\partial \bar{u}_x}{\partial r} + \bar{u}_x \frac{\partial \bar{u}_x}{\partial x } + \frac{1}{\rho} \frac{\partial \bar{p}}{\partial x} - \nu_T \left[\frac{\partial^2\bar{u}_x}{\partial r^2} + \frac{1}{r}\frac{\partial \bar{u}_x}{\partial r} + \frac{\partial^2 \bar{u}_x}{\partial x^2}  \right] -
    2\frac{\partial \nu_T}{\partial x} \frac{\partial \bar{u}_x}{\partial x} - \frac{\partial \nu_T}{\partial r} \left[\frac{\partial \bar{u}_r}{\partial x} + \frac{\partial \bar{u}_x}{\partial r} \right]}_{\mathcal{N}_x(\bar{u}_x,\bar{u}_r,\bar{p},\nu)} \label{eq:swirl_mom1}\\
    0 =& \underbrace{\bar{u}_r \frac{\partial \bar{u}_r}{\partial r}  - \frac{\bar{u}_\theta^2}{r} + \bar{u}_x \frac{\partial \bar{u}_r}{\partial x } + \frac{1}{\rho} \frac{\partial \bar{p}}{\partial r} - \nu_T \left[\frac{\partial^2\bar{u}_r}{\partial r^2} + \frac{1}{r}\frac{\partial \bar{u}_r}{\partial r} - \frac{\bar{u}_r}{r^2} + \frac{\partial^2 \bar{u}_r}{\partial x^2}  \right] -
    2\frac{\partial \nu_T}{\partial r} \frac{\partial \bar{u}_r}{\partial r} - \frac{\partial \nu_T}{\partial x} \left[\frac{\partial \bar{u}_r}{\partial x} + \frac{\partial \bar{u}_x}{\partial r} \right] }_{\mathcal{N}_r(\bar{u}_x,\bar{u}_r,\bar{u}_\theta,\bar{p},\nu)} \\
    0 =& \underbrace{\bar{u}_r \frac{\partial \bar{u}_\theta}{\partial r} + \frac{\bar{u}_r \bar{u}_\theta}{r} + \bar{u}_x \frac{\partial \bar{u}_\theta}{\partial x } - \nu_T \left[\frac{\partial^2\bar{u}_\theta}{\partial r^2} + \frac{1}{r}\frac{\partial \bar{u}_\theta}{\partial r} - \frac{\bar{u}_\theta}{r^2} + \frac{\partial^2 \bar{u}_\theta}{\partial x^2}  \right] -
    \frac{\partial \nu_T}{\partial x} \frac{\partial \bar{u}_\theta}{\partial x} - \frac{\partial \nu_T}{\partial r} \left[\frac{\partial \bar{u}_\theta}{\partial r} - \frac{ \bar{u}_\theta}{r} \right]}_{\mathcal{N}_\theta(\bar{u}_r,\bar{u}_\theta,\nu)} \\
    0 =& \underbrace{\frac{\partial \bar{u}_x}{\partial x} + \frac{\partial \bar{u}_r}{\partial r} + \frac{\bar{u}_r}{r}}_{\mathcal{N}_c(\bar{u}_x, \bar{u}_r)}, \label{eq:swirl_mass}
\end{align}

\begin{acknowledgments}
The authors gratefully acknowledge the help of Henning Bockhorn, Thorsten Zirwes and Feichi Zhang from Karlsruhe Institute of Technology for providing the data of the reacting jet flow. We also want to thank Adrian Hill for his support with Tensorflow.
\end{acknowledgments}


%
%

%


\bibliography{bib.bib}

\end{document}